\documentclass[aps,pra,twocolumn,showpacs,groupedaddress]{revtex4-1}
\usepackage[english]{babel}
\usepackage[T1]{fontenc}
\usepackage{natbib}
\usepackage{amssymb}
\usepackage{bm}
\usepackage{amsmath}
\usepackage[mathcal]{eucal}
\usepackage{color}
\usepackage{graphicx}
\usepackage{hyperref}
\usepackage{mathtools}

\newcommand {\nc} {\newcommand}
\nc {\beq} {\begin{eqnarray}}
\nc {\eeqn} [1] {\label{#1} \end{eqnarray}}
\nc {\eoln} [1] {\label{#1} \\}
\nc {\eol} {\nonumber \\}
\nc {\rref} [1] {(\ref{#1})}
\nc {\Eq} [1] {Eq.~(\ref{#1})}
\nc {\Ref} [1] {Ref.~\cite{#1}}
\nc {\la} {\mbox{$\langle$}}
\nc {\ra} {\mbox{$\rangle$}}
\nc {\dem} {\mbox{$\frac{1}{2}$}}
\nc {\cP} {\mathcal{P}}
\nc {\cN} {\mathcal{N}}
\nc {\ve} [1] {\mbox{\boldmath $#1$}}
\nc {\arrow} [2] {\mbox{$\mathop{\rightarrow}\limits_{#1 \rightarrow #2}$}}
\nc {\red}[1] {\textcolor{red}{#1}}

\newcommand{\bs}[1]{\boldsymbol{#1}}

\DeclarePairedDelimiter\floor{\lfloor}{\rfloor}

\DeclareMathOperator*{\SumInt}{%
\mathchoice%
  {\ooalign{$\displaystyle\sum$\cr\hidewidth$\displaystyle\int$\hidewidth\cr}}
  {\ooalign{\raisebox{.14\height}{\scalebox{.7}{$\textstyle\sum$}}\cr\hidewidth$\textstyle\int$\hidewidth\cr}}
  {\ooalign{\raisebox{.2\height}{\scalebox{.6}{$\scriptstyle\sum$}}\cr$\scriptstyle\int$\cr}}
  {\ooalign{\raisebox{.2\height}{\scalebox{.6}{$\scriptstyle\sum$}}\cr$\scriptstyle\int$\cr}}
}

\begin{document}

\title{Relativistic two-photon decay rates with the Lagrange-mesh method}
\author{Livio Filippin}
\email[]{Livio.Filippin@ulb.ac.be}
\affiliation{Chimie Quantique et Photophysique, C.P.\ 160/09, Universit\'e Libre de Bruxelles (ULB), B-1050 Brussels, Belgium}
\author{Michel Godefroid}
\email[]{mrgodef@ulb.ac.be}
\affiliation{Chimie Quantique et Photophysique, C.P.\ 160/09, Universit\'e Libre de Bruxelles (ULB), B-1050 Brussels, Belgium}
\author{Daniel Baye}
\email[]{dbaye@ulb.ac.be}
\affiliation{Physique Quantique, and Physique Nucl\'{e}aire Th\'{e}orique et Physique Math\'{e}matique, C.P.\ 229, Universit\'e Libre de Bruxelles (ULB), B-1050 Brussels, Belgium}
\date{\today}

\begin{abstract}
Relativistic two-photon decay rates of the $2s_{1/2}$ and $2p_{1/2}$ states towards the $1s_{1/2}$ ground state of hydrogenic atoms are calculated by using numerically exact energies and wave functions obtained from the Dirac equation with the Lagrange-mesh method. This approach is an approximate variational method taking the form of equations on a grid because of the use of a Gauss quadrature approximation. Highly accurate values are obtained by a simple calculation involving different meshes for the initial, final and intermediate wave functions and for the calculation of matrix elements. The accuracy of the results with a Coulomb potential is improved by several orders of magnitude in comparison with benchmark values of the literature. The general requirement of gauge invariance is also successfully tested, down to rounding errors. The method provides high accuracies for two-photon decay rates of a particle in other potentials and is applied to a hydrogen atom embedded in a Debye plasma simulated by a Yukawa potential.
\end{abstract}


\maketitle

%%%%%%%%%%%%%%%%%%%%%%%%%%%%%%%%%%%%%%%%%%%%%%%%%%%%%%%%%%%%%%%%%%%%%%%%%%%%%%%%%%%%%%%%%%%%%%%%%%%%%%%%%%%%%%%%%%%%%%%%%%%%%%%%%%%%%%%%%%%%%%%%%%%%%%%%%%%%%%%%%%%%%%%%%%%%%%%%%%%%%%%%%%%%%%%%%%%%%%%%%%%%%%%%%%%%%%%%%%%%%%%%%%%%%%%%%%%%%%%%%%%%
%%%%%%%%%%%%%%%%%%%%%%%%%%%%%%%%%%%%%%%%%%%%%%%%%%%%%%%%%%%%%%%%%%%%%%%%%%%%%%%%%%%%%%%%%%%%%%%%%%%%%%%%%%%%%%%%%%%%%%%%%%%%%%%%%%%%%%%%%%%%%%%%%%%%%%%%%%%%%%%%%%%%%%%%%%%%%%%%%%%%%%%%%%%%%%%%%%%%%%%%%%%%%%%%%%%%%%%%%%%%%%%%%%%%%%%%%%%%%%%%%%%%

\section{Introduction}
\label{sec:intro}

The radiative decay of the metastable $2s_{1/2}$ state to the $1s_{1/2}$ ground state of hydrogenic atoms is one of the most widely studied atomic transitions in which selection rules forbid the emission of an electric dipole ($E1$) photon. The $2s_{1/2}$ state can decay by two competing processes: the emission of a single photon in a magnetic dipole ($M1$) transition, or the emission of two photons via intermediate virtual states. An important distinction between both processes lies in the fact that, unlike the spectrum in a one-photon process, the emission spectrum of spontaneous two-photon transitions is continuous because energy conservation only requires that the sum of both photon energies equals the transition energy.

Because of its physical importance, the $2s_{1/2} \rightarrow 1s_{1/2}$ two-photon transition rate in hydrogen has been calculated and discussed many times using different approaches. The earliest theoretical work on two-photon processes was performed by G\"{o}ppert-Mayer~\cite{Go31}, who concluded that the simultaneous emission of two electric dipole ($2E1$) photons was the dominant decay mechanism to the $1s$ ground state for the metastable $2s$ state of hydrogen. This conclusion was confirmed by the work of Breit and Teller~\cite{BT40}, who estimated both non-relativistic $2E1$ and $M1$ $2s \rightarrow 1s$ transition rates in hydrogen. They deduced that the dominant two-photon transition is the principal cause of the radiative decay of interstellar $2s$ hydrogen atoms. The earliest interest in these transitions from metastable states of hydrogen came mainly from astrophysics~\cite{SG51,SSS99}, and was recently revived by Chluba and Sunyaev~\cite{CS06}. A historical overview from both theoretical and experimental points of view can be found in \Ref{SPI98}.

By contrast, the interest in the two-photon $2p_{1/2} \rightarrow 1s_{1/2}$ transition is only academic, since the $2p_{1/2}$ state of hydrogenic atoms dominantly decays via an allowed one-photon transition ($E1$), much more probable than competing two-photon processes. 

Usually, the properties of two-photon atomic transitions are evaluated within the framework of the second-order perturbation theory. This calculation involves an infinite number of intermediate virtual states. However, excellent results can be obtained with a finite number of pseudostates, built on complete basis sets such as Dirac Green's functions~\cite{GD81}, $B$ splines~\cite{SPI98}, $B$ polynomials~\cite{ASP11} and Sturmians~\cite{BTE14}. This calculation is simplified by the Lagrange-mesh method, which is an approximate variational method involving a basis of Lagrange functions related to a set of mesh points associated with a Gauss quadrature \cite{BH86,VMB93,Ba15}. Lagrange functions are continuous functions that vanish at all points of the corresponding mesh but one. The principal simplification appearing in the Lagrange-mesh method is that matrix elements are calculated with the associated Gauss quadrature. The potential matrix is then diagonal and only involves values of the potential at mesh points. Recently, we have shown that numerically exact solutions of the Coulomb-Dirac equation can be obtained with this method \cite{Ba15,BFG14}. More generally, the method is very accurate for most central potentials as illustrated with Yukawa potentials in \Ref{BFG14}. Hence, it allows the calculation of two-photon decay rates in various types of potentials. Here we study the $2s_{1/2} \rightarrow 1s_{1/2}$ and $2p_{1/2} \rightarrow 1s_{1/2}$ transitions in the hydrogenic and Yukawa cases.

In Sec.~\ref{sec:formulation}, the relativistic expressions of two-photon decay rates of an electron in a potential are recalled. In Sec.~\ref{sec:LMM}, the principle of the Lagrange-mesh method is summarized and relativistic two-photon decay rates are approximated with Gauss quadratures. In Sec.~\ref{sec:res}, numerical results are presented for hydrogenic atoms and for a particle in Yukawa potentials. Section \ref{sec:conc} contains concluding remarks.

Except in \tablename{~\ref{table_2s_1s_Coulomb}, we use for the fine-structure constant and the atomic unit of time the 2010 CODATA recommended values $1/\alpha=137.035\,999\,074$ and $\hbar/E_h=2.418\,884\,326\,502 \times 10^{-17}$ s~\cite{MTN12}, for the sake of comparison with previous works.

%%%%%%%%%%%%%%%%%%%%%%%%%%%%%%%%%%%%%%%%%%%%%%%%%%%%%%%%%%%%%%%%%%%%%%%%%%%%%%%%%%%%%%%%%%%%%%%%%%%%%%%%%%%%%%%%%%%%%%%%%%%%%%%%%%%%%%%%%%%%%%%%%%%%%%%%%%%%%%%%%%%%%%%%%%%%%%%%%%%%%%%%%%%%%%%%%%%%%%%%%%%%%%%%%%%%%%%%%%%%%%%%%%%%%%%%%%%%%%%%%%%%
%%%%%%%%%%%%%%%%%%%%%%%%%%%%%%%%%%%%%%%%%%%%%%%%%%%%%%%%%%%%%%%%%%%%%%%%%%%%%%%%%%%%%%%%%%%%%%%%%%%%%%%%%%%%%%%%%%%%%%%%%%%%%%%%%%%%%%%%%%%%%%%%%%%%%%%%%%%%%%%%%%%%%%%%%%%%%%%%%%%%%%%%%%%%%%%%%%%%%%%%%%%%%%%%%%%%%%%%%%%%%%%%%%%%%%%%%%%%%%%%%%%%

\section{Relativistic formulation}
\label{sec:formulation}

In atomic units $\hbar=m_e=e=1$ where $m_e$ is the electron mass, the Dirac Hamiltonian reads \cite{Gr07}
\beq
H_D = c \ve{\alpha}\cdot\ve{p} + \beta c^2 + V(r)
\eeqn{dir.1}
where $c=1/\alpha$ is the speed of light, $\ve{p}$ is the momentum operator, $V$ is the potential and $\ve{\alpha}$ and $\beta$ are the traditional Dirac matrices. The eigenenergies of $H_D$ are denoted as $c^2+E$ and the Dirac equation reads  
\beq
H_D \; \phi_{n\kappa m} (\ve{r}) = (c^2 + E) \;  \phi_{n\kappa m} (\ve{r}).
\eeqn{dir.0}
The Dirac spinors are defined as 
\beq
\phi_{n\kappa m} (\ve{r}) = \frac{1}{r} \left( \begin{array}{c} 
P_{n\kappa}(r) \chi_{\kappa m} \\ iQ_{n\kappa}(r) \chi_{-\kappa m} \end{array} \right)
\eeqn{dir.2}
as a function of the large and small radial components, $P_{n\kappa}(r)$ and $Q_{n\kappa}(r)$ respectively. The quantum number $n$ labels the different states with the same symmetry. The spinors $\chi_{\kappa m}$ are common eigenstates of $\ve{L}^2$, $\ve{S}^2$, $\ve{J}^2$, and $J_z$ with respective eigenvalues $l(l+1)$, 3/4, $j(j+1)$, and $m$ where 
\beq
j = |\kappa| - \dem, \quad l = j + \dem\, \mathrm{sgn}\, \kappa.
\eeqn{dir.3}
The coupled radial Dirac equations read in matrix form
\beq
H_\kappa \left( \begin{array}{c} P_{n\kappa}(r) \\ Q_{n\kappa}(r) \end{array} \right) = E_{n\kappa} \left( \begin{array}{c} P_{n\kappa}(r) \\ Q_{n\kappa}(r) \end{array} \right)
\eeqn{r.1}
with the Hamiltonian matrix 
\beq
H_\kappa = \left( \begin{array}{c c} V(r) & c \left( -\frac{d}{dr} + \frac{\kappa}{r} \right) \\ c \left( \frac{d}{dr} + \frac{\kappa}{r} \right) & V(r) - 2c^2 \end{array} \right),
\eeqn{r.2}

The large and small radial functions, $P_{n\kappa}(r)$ and $Q_{n\kappa}(r)$, are normalized according to the condition $\int_0^\infty \{[P_{n\kappa}(r)]^2+[Q_{n\kappa}(r)]^2\}dr$\,=\,1. If $V_0=-\lim_{r \rightarrow 0}rV(r)$ is strictly positive, and $V_0<c$, they behave at the origin as~\cite{Gr07}
\beq
P_{n\kappa}(r),\ Q_{n\kappa}(r) \arrow{r}{0} r^\gamma,
\eeqn{r.10}
where the parameter $\gamma$ is defined by
\beq
\gamma = \sqrt{\kappa^2 - (V_0/c)^2}.
\eeqn{r.11}
The Dirac spinors are singular at the origin for $\gamma<1$. This singularity can be important for hydrogenic ions with high nuclear charges $Z$. 
In the $V_0 = 0$ case, $P_{n\kappa}$ and $Q_{n\kappa}$ behave at the origin as 
$r^{|\kappa|}$ or $r^{|\kappa|+1}$ ~\cite{Gr07}.

In the Coulomb case, the potential is $V(r)=-Z/r$ in atomic units. Constant $V_0$ is equal to $Z$. The energy of level $n\kappa$ is
\beq
E_{n \kappa} = - \dfrac{Z^2}{\cN (\cN + n - |\kappa| + \gamma)},
\eeqn{c.1}
with
\beq
\cN = [(n - |\kappa| + \gamma)^2 + (\alpha Z)^2]^{1/2}. 
\eeqn{c.2}

For a system described with the Dirac equation, the basic expression for the differential decay rate (in energy of one of the photons) reads, in atomic units~\cite{GD81},
\beq
\dfrac{dw}{d\omega_1} \, d\Omega_1 d\Omega_2 & = & \dfrac{\omega_1 \omega_2}{8\pi^3c^2} \left| \SumInt_\nu \left( \dfrac{\langle f \vert A_2^\ast \vert \nu \rangle \langle \nu \vert A_1^\ast \vert i \rangle}{E_\nu-E_i+\omega_1} \right. \right. \eol
& & + \left. \left. \dfrac{\langle f \vert A_1^\ast \vert \nu \rangle \langle \nu \vert A_2^\ast \vert i \rangle}{E_\nu-E_i+\omega_2} \right) \right|^2 d\Omega_1 d\Omega_2,
\eeqn{eq_dw}
where $\vert i \rangle \equiv \vert n_i \kappa_i m_i \rangle$ and $\vert f \rangle \equiv \vert n_f \kappa_f m_f \rangle$ correspond, in $\vert \bs{r} \rangle$ representation, to Dirac spinors \rref{dir.2} of the initial and final states with respective energies $E_i$ and $E_f$, $\omega_j$ is the frequency and $d\Omega_j$ is the element of solid angle for the $j$th photon. The transition proceeds through an infinite set of intermediate states $\vert \nu \rangle \equiv \vert n_\nu \kappa_\nu m_\nu \rangle$ at energy $E_\nu$. The summation over $\nu$ includes integrations over the continua for both positive and negative energy solutions of the Dirac equation. The frequencies of the photons are constrained by energy conservation
\beq
E_i - E_f = \omega_1 + \omega_2,
\eeqn{conserv_energy}
where the recoil of the nucleus is neglected.

For a photon plane wave with propagation vector $\bs{k}_j$ and polarization vector $\hat{\bs{e}}_j$ $(\hat{\bs{e}}_j \cdot \bs{k}_j=0)$, the operators $A_j^\ast$ in \Eq{eq_dw} are given by
\beq
A_j^\ast = \bs{\alpha} \cdot (\hat{\bs{e}}_j + G \hat{\bs{k}}_j) e^{-i \bs{k}_j \cdot \bs{r}} - G e^{-i \bs{k}_j \cdot \bs{r}},
\eeqn{eq_A_j}
where $G$ is an arbitrary gauge parameter controlling the contribution from fictitious longitudinal and scalar photon states~\cite{GD81}. Among the large variety of possible gauges, Grant~\cite{Gr74} showed that two values of $G$ are of particular interest since they lead to well-known non-relativistic operators. The $G=0$ value defines the so-called Coulomb gauge, or velocity gauge, which leads to the electric dipole velocity form in the non-relativistic limit. The value $G=[(L+1)/L]^{1/2}$, where $L$ is the multipolarity, defines the so-called Babushkin gauge, or length gauge, which leads to the non-relativistic electric dipole length form of the transition operator. From the general requirement of gauge invariance, the final results must be independent of $G$. The gauge invariance of the two-photon relativistic calculations was studied by Goldman and Drake~\cite{GD81}, Santos \textit{et al.}~\cite{SPI98} and Amaro \textit{et al.}~\cite{ASP11}.

Let us denote by $dW/d\omega_1$ the differential decay rate \rref{eq_dw} summed over the transverse polarizations $\hat{\bs{e}}_1$, $\hat{\bs{e}}_2$ and integrated over $d\Omega_1$, $d\Omega_2$, i.e.,
\beq
\frac{dW}{d\omega_1} = \sum_{\hat{\bs{e}}_1,\hat{\bs{e}}_2} \int \int \frac{dw}{d\omega_1} \, d\Omega_1 d\Omega_2.
\eeqn{eq_dW}
The average partial decay rates, i.e., summed over the magnetic quantum number $m_f$ and averaged over $m_i$, describing the two-photon transitions of a given type $\lambda$ and multipolarity $L$ are given by~\cite{GD81}
\beq
& & \dfrac{d\overline{W}_{\lambda_1 L_1,\lambda_2 L_2}}{d\omega_1} = \dfrac{\omega_1 \omega_2}{8\pi^3c^2(2j_i+1)} \eol
& & \times \sum_{j_\nu} \left\lbrace \left[ S_{\lambda_1 L_1,\lambda_2 L_2}^{j_\nu}(2,1) \right]^2 + \left[ S_{\lambda_1 L_1,\lambda_2 L_2}^{j_\nu}(1,2) \right]^2 \right. \eol
& & \,\,\, + 2\sum_{j'_\nu} \left. d_{L_1,L_2}^{j_\nu,j'_\nu} S_{\lambda_1 L_1,\lambda_2 L_2}^{j_\nu}(2,1) S_{\lambda_1 L_1,\lambda_2 L_2}^{j'_\nu}(1,2) \right\rbrace.
\eeqn{eq_dWL1lam1L2lam2}
$L$ is the photon angular momentum and $\lambda$ stands for the electric ($\lambda=1$), magnetic ($\lambda=0$) and longitudinal ($\lambda=-1$) terms. The factor $d_{L_1,L_2}^{j_\nu,j'_\nu}$ involves a 6-$j$ symbol representing the angular couplings,
\beq
d_{L_1,L_2}^{j_\nu,j'_\nu} = (-1)^{2j'_\nu+L_1+L_2}[j_\nu,j'_\nu]^{1/2} \begin{Bmatrix} j_f & j'_\nu & L_1 \\ j_i & j_\nu & L_2 \end{Bmatrix},
\eeqn{eq_djjp}
and $S_{\lambda_1 L_1,\lambda_2 L_2}^{j_\nu}(2,1)$ reads
\beq
& & S_{\lambda_1 L_1,\lambda_2 L_2}^{j_\nu}(2,1) = \Delta_{\lambda_1 L_1,\lambda_2 L_2}^{j_\nu}(2,1) \eol
& & \qquad \qquad \times \sum_{\kappa_\nu} \SumInt_{n_\nu} \dfrac{\overline{M}_{f,n_\nu \kappa_\nu}^{(\lambda_2,L_2)}(\omega_2) \, \overline{M}_{n_\nu \kappa_\nu,i}^{(\lambda_1,L_1)}(\omega_1)}{E_{n_\nu \kappa_\nu}-E_i+\omega_1},
\eeqn{eq_Sj21}
where
\beq
\Delta_{\lambda_1 L_1,\lambda_2 L_2}^{j_\nu}(2,1) & = & \frac{4\pi [j_i,j_\nu,j_f]^{1/2}}{[L_1,L_2]^{1/2}} \begin{pmatrix} j_f & L_2 & j_\nu \\ 1/2 & 0 & -1/2 \end{pmatrix} \eol
& \,\,\,\, \times & \begin{pmatrix} j_\nu & L_1 & j_i \\ 1/2 & 0 & -1/2 \end{pmatrix} \pi_{i \nu}^{\lambda_1 L_1} \pi_{\nu f}^{\lambda_2 L_2}.
\eeqn{eq_Deltaj21}
$S_{\lambda_1 L_1,\lambda_2 L_2}^{j_\nu}(1,2)$ is analogously obtained by permuting indices 1 and 2. The notation $[a,b,...]$ in Eqs.~\rref{eq_djjp} and \rref{eq_Deltaj21} means $(2a+1)(2b+1)\cdots$. The factors $\pi_{i \nu}^{\lambda_1 L_1}$ and $\pi_{\nu f}^{\lambda_2 L_2}$ are given by
\beq
\pi_{\alpha \beta}^{\lambda L} = \left\lbrace \begin{aligned} 1 \quad & \text{if} \quad l_\alpha + l_\beta + \lambda + L \,\, \text{odd}, \\ 0 \quad & \text{if} \quad l_\alpha + l_\beta + \lambda + L \,\, \text{even}. \end{aligned} \right.
\eeqn{eq_pi1pi2}
Parity selection rules \rref{eq_pi1pi2} follow from the calculation of the matrix elements appearing in \Eq{eq_dw}. As the sum over $\nu$ in \Eq{eq_dw}, the sum over $n_\nu$ in \Eq{eq_Sj21} represents a sum over discrete states and an integral over the continuum involving also negative energy states.

The radial matrix elements $\overline{M}_{\alpha,\beta}^{(\lambda,L)}$ appearing in \Eq{eq_Sj21} are given by~\cite{Gr74}
\beq
\overline{M}_{\alpha,\beta}^{(1,L)} & = & \left( \frac{L}{L+1} \right)^{1/2} \left[ (\kappa_\alpha-\kappa_\beta) I_{L+1}^{+} + (L+1) I_{L+1}^{-} \right] \eol
& & - \left( \frac{L+1}{L} \right)^{1/2} \left[ (\kappa_\alpha-\kappa_\beta) I_{L-1}^{+} - L I_{L-1}^{-} \right], \eol
\overline{M}_{\alpha,\beta}^{(0,L)} & = & \frac{2L+1}{[L(L+1)]^{1/2}} \, (\kappa_\alpha+\kappa_\beta) I_{L}^{+}, \eol
\overline{M}_{\alpha,\beta}^{(-1,L)} & = & G\left[ (2L+1) J_L + (\kappa_\alpha-\kappa_\beta) (I_{L+1}^{+}+I_{L-1}^{+}) \right. \eol
& & \quad - \left. L I_{L-1}^{-} + (L+1) I_{L+1}^{-} \right],
\eeqn{eq_Mfi}
with the radial integrals $I_L^{\pm}$ and $J_L$ defined following the notation by Rosner and Bhalla~\cite{RB70} 
\beq
I_L^{\pm}(\omega) = \int_0^\infty (P_\alpha Q_\beta \pm Q_\alpha P_\beta) \, j_L\left( \frac{\omega}{c}r \right) dr
\eeqn{eq_ILpm}
and
\beq
J_L(\omega) = \int_0^\infty (P_\alpha P_\beta + Q_\alpha Q_\beta) \, j_L\left( \frac{\omega}{c}r \right) dr,
\eeqn{eq_JL}
where $j_L(x)$ is a spherical Bessel function of the first kind~\cite{AS65}.

The total decay rate for a transition in which one $\Theta_1 L_1$ photon and one $\Theta_2 L_2$ photon are emitted, where $\Theta_i=E,M$ stand for the electric and magnetic multipole types, is given by
\beq
\overline{W}_{\Theta_1 L_1 \Theta_2 L_2} = \sum_{\lambda_{\Theta_1}, \lambda_{\Theta_2}} \int_{0}^{\omega_t} \dfrac{d\overline{W}_{\lambda_{\Theta_1} L_1,\lambda_{\Theta_2} L_2}}{d\omega_1} \, d\omega_1,
\eeqn{eq_WTh1L1Th2L2}
where $\omega_t$ is the energy of the two-photon transition, $\omega_t=\omega_1+\omega_2=E_i-E_f$, and
\beq
\lambda_{\Theta_i} = \left\lbrace \begin{aligned} 1,-1 \quad & \text{if} \quad \Theta_i=E, \\ 0 \quad & \text{if} \quad \Theta_i=M. \end{aligned} \right.
\eeqn{eq_lam_Th_i}

Equation \rref{eq_dWL1lam1L2lam2} for the decay rate, which depends quadratically on $G$~\cite{GD81}, is called \textit{incoherent} in the literature~\cite{SSA09}. However, a gauge transformation~\cite{Gr74} gives rise to the electric $E$ and magnetic $M$ matrix elements
\beq
\overline{\mathcal{M}}_{\alpha,\beta}^{(E,L)} & = & \overline{M}_{\alpha,\beta}^{(1,L)} + \overline{M}_{\alpha,\beta}^{(-1,L)}, \eol
\overline{\mathcal{M}}_{\alpha,\beta}^{(M,L)} & = & \overline{M}_{\alpha,\beta}^{(0,L)},
\eeqn{eq_gauge_transfo}
and uses the $\overline{\mathcal{M}}$'s in place of the $\overline{M}$'s in \Eq{eq_Sj21}. Since $\overline{M}_{\alpha,\beta}^{(-1,L)}$ is linear in $G$, the final results also contain terms linear in $G$~\cite{GD81}. Applying \rref{eq_gauge_transfo} to the reduced matrix elements \rref{eq_Sj21} for electric type $E$ of transition, $S_{E L_1,E L_2}^{j_\nu}(2,1)$, leads to
\beq
& & S_{E L_1,E L_2}^{j_\nu}(2,1) = \Delta_{\lambda_1 L_1,\lambda_2 L_2}^{j_\nu}(2,1) \eol
& & \quad \times \sum_{\kappa_\nu} \SumInt_{n_\nu} \dfrac{1}{E_{n_\nu \kappa_\nu}-E_i+\omega_1} \eol
& & \quad \quad \quad \quad \quad \times \left[ \overline{M}_{f,n_\nu \kappa_\nu}^{(1,L_2)}(\omega_2) + \overline{M}_{f,n_\nu \kappa_\nu}^{(-1,L_2)}(\omega_2) \right] \eol
& & \quad \quad \quad \quad \quad \times \left[ \overline{M}_{n_\nu \kappa_\nu,i}^{(1,L_1)}(\omega_1) + \overline{M}_{n_\nu \kappa_\nu,i}^{(-1,L_1)}(\omega_1) \right] \eol
& & = S_{1 L_1,1 L_2}^{j_\nu}(2,1) + S_{1 L_1,-1 L_2}^{j_\nu}(2,1) \eol
& & \quad + S_{-1 L_1,1 L_2}^{j_\nu}(2,1) + S_{-1 L_1,-1 L_2}^{j_\nu}(2,1).
\eeqn{eq_coherent}
The same argument holds for $S_{M L_1,E L_2}^{j_\nu}(2,1)$, leading to $S_{0 L_1,1 L_2}^{j_\nu}(2,1)+S_{0 L_1,-1 L_2}^{j_\nu}(2,1)$. These summations are referred as \textit{coherent} in the literature~\cite{SSA09}. Hence, $\overline{W}_{\Theta_1 L_1 \Theta_2 L_2}$ simply reads
\beq
\overline{W}_{\Theta_1 L_1 \Theta_2 L_2} = \int_{0}^{\omega_t} \dfrac{d\overline{W}_{\lambda_{\Theta_1} L_1,\lambda_{\Theta_2} L_2}}{d\omega_1} \, d\omega_1,
\eeqn{eq_WTh1L1Th2L2_coherent}
with
\beq
\lambda_{\Theta_i} = \left\lbrace \begin{aligned} 1 \quad & \text{if} \quad \Theta_i=E, \\ 0 \quad & \text{if} \quad \Theta_i=M. \end{aligned} \right.
\eeqn{eq_lam_Th_i_coher}
It has been proven in the Appendix of \Ref{GD81} that the exact decay rates remain independent of $G$. The incoherent and coherent summations are compared in Sec.~\ref{sec:res}, to test the accuracy of the numerical results.

Finally, for both incoherent and coherent summations, the total spontaneous emission probability per unit of time for a two-photon transition is obtained by summing over all allowed multipole components,
\beq
W = \sum_{\text{all} \, \Theta_1 L_1,\Theta_2 L_2} \, t_{\Theta_1 L_1,\Theta_2 L_2} \, \overline{W}_{\Theta_1 L_1 \Theta_2 L_2},
\eeqn{eq_W}
where
\beq
t_{\Theta_1 L_1,\Theta_2 L_2} = \left\lbrace \begin{aligned} 1 \quad & \text{if} \quad \Theta_1 L_1 \neq \Theta_2 L_2, \\ 1/2 \quad & \text{if} \quad \Theta_1 L_1 = \Theta_2 L_2. \end{aligned} \right.
\eeqn{eq_t}
The factor of 1/2 is included to avoid counting twice each pair, when both photons have the same characteristics.

For further use in Sec.~\ref{sec:res}, the spontaneous emission rate for a one-photon transition $i \rightarrow f$ is \cite{Gr74}, in atomic units,
\beq
W_{i \rightarrow f} = 2 \alpha \omega \frac{[j_f]}{[L]} \pi_{fi}^{\lambda L} \begin{pmatrix} j_i & L & j_f \\ 1/2 & 0 & -1/2 \end{pmatrix}^2 \vert \overline{\mathcal{M}}_{fi} \vert^2,
\eeqn{eq_one_photon}
where $\pi_{fi}^{\lambda L}$ and $\overline{\mathcal{M}}_{fi}$ are respectively given by Eqs.~\rref{eq_pi1pi2} and \rref{eq_gauge_transfo}, and $\omega=E_i-E_f$ is the transition energy.

%%%%%%%%%%%%%%%%%%%%%%%%%%%%%%%%%%%%%%%%%%%%%%%%%%%%%%%%%%%%%%%%%%%%%%%%%%%%%%%%%%%%%%%%%%%%%%%%%%%%%%%%%%%%%%%%%%%%%%%%%%%%%%%%%%%%%%%%%%%%%%%%%%%%%%%%%%%%%%%%%%%%%%%%%%%%%%%%%%%%%%%%%%%%%%%%%%%%%%%%%%%%%%%%%%%%%%%%%%%%%%%%%%%%%%%%%%%%%%%%%%%%
%%%%%%%%%%%%%%%%%%%%%%%%%%%%%%%%%%%%%%%%%%%%%%%%%%%%%%%%%%%%%%%%%%%%%%%%%%%%%%%%%%%%%%%%%%%%%%%%%%%%%%%%%%%%%%%%%%%%%%%%%%%%%%%%%%%%%%%%%%%%%%%%%%%%%%%%%%%%%%%%%%%%%%%%%%%%%%%%%%%%%%%%%%%%%%%%%%%%%%%%%%%%%%%%%%%%%%%%%%%%%%%%%%%%%%%%%%%%%%%%%%%%

\section{Lagrange-mesh method}
\label{sec:LMM}

\subsection{Mesh equations}
\label{subsec:mesh}

The principles of the Lagrange-mesh method are described in Refs.~\cite{BH86,VMB93,Ba15} and its application to the Dirac equation is presented in Refs.~\cite{Ba15,BFG14}. The mesh points $x_j$ are defined by~\cite{BH86} 
\beq
L_N^{\alpha}(x_j) = 0,
\eeqn{Lag.1}
where $j=1$ to $N$ and $L_N^{\alpha}$ is a generalized Laguerre polynomial depending on parameter $\alpha$~\cite{AS65}. This mesh is associated with a Gauss-Laguerre quadrature 
\beq
\int_0^\infty g(x) \, dx \approx \sum^N_{k=1} \lambda_k \, g(x_k), 
\eeqn{Lag.2}
with the weights $\lambda_k$. The Gauss quadrature is exact for the Laguerre weight function $x^{\alpha}e^{-x}$ multiplied by any polynomial of degree at most $2N-1$~\cite{Sz67}. 

The regularized Lagrange functions are defined by~\cite{Ba95,BHV02,Ba15}
\beq
\hat{f}_j^{(\alpha)} (x) & = & (-1)^j \left( \frac{N!}{\Gamma (N + \alpha + 1) x_j} \right)^{1/2} \eol
& & \times \frac{L_N^{\alpha}(x)}{x-x_j} x^{\alpha/2+1} e^{-x/2}.
\eeqn{Lag.3}
The functions $\hat{f}_j^{(\alpha)}(x)$ are polynomials of degree $N-1$ multiplied by $x$ and by the square root of the Laguerre weight $x^{\alpha}\exp(-x)$. The Lagrange functions satisfy the Lagrange conditions 
\beq
\hat{f}_j^{(\alpha)}(x_i) = \lambda_i^{-1/2} \delta_{ij}.
\eeqn{Lag.4}
They are not orthonormal, but become orthonormal at the Gauss-quadrature approximation. Condition \rref{Lag.4} drastically simplifies the expressions calculated with the Gauss quadrature. 

The radial functions $P_{n\kappa}(r)$ and $Q_{n\kappa}(r)$ are expanded in regularized Lagrange functions \rref{Lag.3} as 
\beq
P_{n\kappa}(r) = h^{-1/2} \sum_{j=1}^{N} \; p_{n\kappa j} \hat{f}_j^{(\alpha)}(r/h),
\eoln{Lag.5}
Q_{n\kappa}(r) = h^{-1/2} \sum_{j=1}^{N} \; q_{n\kappa j} \hat{f}_j^{(\alpha)}(r/h),
\eeqn{Lag.6}
where $h$ is a scaling parameter aimed at adapting the mesh points $hx_i$ to the physical extension of the problem and $\sum_{j=1}^N \left( p_{n\kappa j}^2+q_{n\kappa j}^2 \right)=1$ ensures the normalization of $P_{n\kappa}(r)$ and~$Q_{n\kappa}(r)$. 

The parameter $\alpha=2(\gamma-1)$ can be selected so that the Lagrange functions behave as $r^\gamma$ near the origin~\cite{BFG14}. Here, another choice $\alpha=2(\gamma-|\kappa|)$ is preferable as explained below. The basis functions then behave as $r^{\gamma-|\kappa|+1}$ but the physical $r^\gamma$ behavior can be simulated by linear combinations. In the Coulomb case, the correct exponential behavior of the components is obtained with $h=\cN/2Z$. Expansions with $N \geq n+|\kappa|$ such functions are able to exactly reproduce the large and small hydrogenic components. 

Let us introduce expansions \rref{Lag.5} and \rref{Lag.6} in the coupled radial Dirac equations \rref{r.1}. Projecting on the Lagrange functions and using the associated Gauss quadrature leads to the $2N \times 2N$ Hamiltonian matrix 
\beq
\ve{H}_\kappa^G = \left( \begin{array}{c c} V(hx_i) \delta_{ij} & \frac{c}{h} \left( D_{ji}^G + \frac{\kappa}{x_i} \delta_{ij} \right) \\ \frac{c}{h} \left( D_{ij}^G + \frac{\kappa}{x_i} \delta_{ij} \right) & (V(hx_i) -2c^2) \delta_{ij} 
\end{array} \right)
\eeqn{Lag.7}
with a $2\times2$ block structure, where 
\beq
D_{i \neq j}^G = (-1)^{i-j} \sqrt{\frac{x_i}{x_j}}\, \frac{1}{x_i-x_j}, \quad D_{ii}^G = \frac{1}{2x_i}.
\eeqn{Lag.8}
Expressions \rref{Lag.8} are the matrix elements $\la \hat{f}_i^{(\alpha)}|d/dx|\hat{f}_j^{(\alpha)} \ra$ calculated at the Gauss-quadrature approximation. This corresponds to choosing the Gauss quadrature `Gauss(2,1)' in \Ref{BFG14}. Notice that the subscripts $i$ and $j$ should be interchanged in Eq.~(24) of \Ref{FGB14}.

In the Coulomb case, an exact variational treatment of the Dirac-Coulomb problem is possible with a Lagrange basis. However, as proven in Appendix~\ref{sec:var_exact}, the Lagrange-mesh equations based on the Hamiltonian matrix \rref{Lag.7} also provide the exact solution of the Dirac-Coulomb problem. If $N \ge n+|\kappa|$ and $h=\cN/2Z$, one of the eigenvalues of $\ve{H}_\kappa^G$ is the exact energy $E_{n\kappa}$ and the corresponding eigenvector $(p_{n\kappa 1},p_{n\kappa 2},\dots,p_{n\kappa N},q_{n\kappa 1},q_{n\kappa 2},\dots,q_{n\kappa N})^T$ provides the coefficients of the exact eigenfunctions in the expansions \rref{Lag.5} and \rref{Lag.6}~\cite{BFG14}. For other potentials, if $N$ is large enough and $h$ well chosen, some negative energies above $-2c^2$ correspond to physical energies. The corresponding eigenvectors provide approximations of the wave functions. 

%%%%%%%%%%%%%%%%%%%%%%%%%%%%%%%%%%%%%%%%%%%%%%%%%%%%%%%%%%%%%%%%%%%%%%%%%%%%%%%%%%%%%%%%%%%%%%%%%%%%%%%%%%%%%%%%%%%%%%%%%%%%%%%%%%%%%%%%%%%%%%%%%%%%%%%%%%%%%%%%%%%%%%%%%%%%%%%%%%%%%%%%%%%%%%%%%%%%%%%%%%%%%%%%%%%%%%%%%%%%%%%%%%%%%%%%%%%%%%%%%%%%

\subsection{Two-photon decay rates on Lagrange meshes}
\label{subsec:two-photon_mesh}

Two-photon decays proceed through an infinite set of intermediate states with some value of $\kappa'$. Finite-basis techniques such as the Lagrange-mesh method allow a discretisation of the continuum, leading to a truncated sum over $2N'$ intermediate states. Some of these states may correspond to exact eigenstates of the Dirac Hamiltonian \rref{dir.1}, while the other ones, discretising the continuum, have no physical meaning and are called \textit{pseudostates}.

Let $E_{n'\kappa'}$, $n'=1,\dots,2N'$, be the eigenvalues of matrix $\ve{H}_{\kappa'}^G$ defined in \Eq{Lag.7} with $\kappa'$ replacing $\kappa$. The corresponding eigenvectors contain the coefficients $p_{n'\kappa'j}$ and $q_{n'\kappa'j}$ of the components $P_{n'\kappa'}$ and $Q_{n'\kappa'}$ of the intermediate states.

The intermediate states are calculated with $\alpha'=2(\gamma'-|\kappa'|)$ in place of $\alpha=2(\gamma-|\kappa|)$. Hence they have the exact behavior $r^{\gamma'}$ at the origin. Matrix $\ve{H}_{\kappa'}^G$ is calculated on a different mesh $h'x'_j$ with $N'$ mesh points, where $h'=fh_f+(1-f)h_i$, $h_i$ and $h_f$ respectively associated to the initial and final states. In the following the value $f=1$ is chosen, i.e., $h'=h_f$. The $f$ value does not influence the numerical results, but can accelerate their convergence. Notice that the values of $\alpha$ and $\alpha'$ are close to each other, much closer than with the choices $\alpha'=2(\gamma'-1)$ and $\alpha=2(\gamma-1)$. The integrand in Eqs.~\rref{eq_ILpm} and \rref{eq_JL} explicitly contains $r^{\gamma+\gamma'}$. In the Coulomb case with $h=\mathcal{N}/2Z$, it is the product of $r^{\gamma+\gamma'}\exp(-2Zr/\cN)$, a polynomial, and a spherical Bessel function. An accurate calculation of Eqs.~\rref{eq_ILpm} and \rref{eq_JL} with a Gauss-Laguerre quadrature is possible by choosing a third mesh $\bar{h}\bar{x}_i$ with $N_{G}$ mesh points, where $\bar{h}=2hh'/(h+h')$, with $h=h_i$ or $h_f$, and $N_{G} \geq \floor{N+N'}/2$. The $\bar{x}_i$ correspond to the weight function $x^{\bar{\alpha}}\exp(-x)$ with the average value
\beq
\bar{\alpha} = \dem (\alpha + \alpha').
\eeqn{Lag.9}
The corresponding weights are denoted as $\bar{\lambda}_i$. 

Let us replace in \Eq{eq_Sj21} the notations $n_\nu$, $j_\nu$ and $\kappa_\nu$ related to the intermediate states $\nu$ by $n'$, $j'$ and $\kappa'$. Approximate radial parts $S_{\lambda_1 L_1,\lambda_2 L_2}^{j'}(2,1)$ are given by
\beq
& & S_{\lambda_1 L_1,\lambda_2 L_2}^{j'}(2,1) = \Delta_{\lambda_1 L_1,\lambda_2 L_2}^{j'}(2,1) \eol
& & \quad \quad \quad \times \sum_{\kappa'} \sum_{n'=1}^{2N'} \dfrac{\overline{M}_{f,n'\kappa'}^{(\lambda_2,L_2)}(\omega_2) \, \overline{M}_{n'\kappa',i}^{(\lambda_1,L_1)}(\omega_1)}{E_{n'\kappa'}-E_i+\omega_1}.
\eeqn{eq_Sj21_Gauss}
$S_{\lambda_1 L_1,\lambda_2 L_2}^{j'}(1,2)$ is analogously obtained. The radial integrals appearing in $\overline{M}_{f,n'\kappa'}^{(\lambda_2,L_2)}(\omega_2)$ and $\overline{M}_{n'\kappa',i}^{(\lambda_1,L_1)}(\omega_1)$ can be obtained from Eqs. \rref{eq_ILpm} and \rref{eq_JL} with the Gauss-Laguerre quadrature~as
\beq
I_L^{\pm}(\omega) \approx \sum_{j=1}^N \sum_{j'=1}^{N'} \left[ p_{n\kappa j} q_{n'\kappa' j'} \pm q_{n\kappa j} p_{n'\kappa' j'} \right] \, \mathcal{J}_{jj'}
\eeqn{Lag.10}
and
\beq
J_L(\omega) \approx \sum_{j=1}^N \sum_{j'=1}^{N'} \left[ p_{n\kappa j} p_{n'\kappa' j'} + q_{n\kappa j} q_{n'\kappa' j'} \right] \, \mathcal{J}_{jj'},
\eeqn{Lag.11}
where 
\beq
\mathcal{J}_{jj'} & = & \int_0^\infty h^{-1/2} \hat{f}_{j}^{(\alpha)}(r/h) j_L(\omega r/c) h'^{-1/2} \hat{f}_{j'}^{(\alpha')}(r/h') \, dr \eol
& \approx & \bar{h} (hh')^{-1/2} \eol
& & \times \sum_{i=1}^{N_G} \bar{\lambda}_{i} \hat{f}_{j}^{(\alpha)}(\bar{h}\bar{x}_i/h) j_L(\omega \bar{h}\bar{x}_{i}/c) \hat{f}_{j'}^{(\alpha')}(\bar{h}\bar{x}_i/h').
\eeqn{Lag.12}
Evaluating integral \rref{Lag.12} requires the explicit computation of Lagrange functions. Some remarks on their numerical calculation can be found in Appendix B of \Ref{FGB14}.

The integral over $\omega_1$ appearing in \Eq{eq_WTh1L1Th2L2} is evaluated with a Gauss-Legendre quadrature involving $N_{\omega_1}$ mesh points.
\clearpage

%%%%%%%%%%%%%%%%%%%%%%%%%%%%%%%%%%%%%%%%%%%%%%%%%%%%%%%%%%%%%%%%%%%%%%%%%%%%%%%%%%%%%%%%%%%%%%%%%%%%%%%%%%%%%%%%%%%%%%%%%%%%%%%%%%%%%%%%%%%%%%%%%%%%%%%%%%%%%%%%%%%%%%%%%%%%%%%%%%%%%%%%%%%%%%%%%%%%%%%%%%%%%%%%%%%%%%%%%%%%%%%%%%%%%%%%%%%%%%%%%%%%
%%%%%%%%%%%%%%%%%%%%%%%%%%%%%%%%%%%%%%%%%%%%%%%%%%%%%%%%%%%%%%%%%%%%%%%%%%%%%%%%%%%%%%%%%%%%%%%%%%%%%%%%%%%%%%%%%%%%%%%%%%%%%%%%%%%%%%%%%%%%%%%%%%%%%%%%%%%%%%%%%%%%%%%%%%%%%%%%%%%%%%%%%%%%%%%%%%%%%%%%%%%%%%%%%%%%%%%%%%%%%%%%%%%%%%%%%%%%%%%%%%%%

\onecolumngrid

\begin{table}[ht!]
\caption{\footnotesize Multipole contributions (in s$^{-1}$) to the $2s_{1/2} \rightarrow 1s_{1/2}$ two-photon decay rate of hydrogenic ions with $Z=1$, 40, and 92. Comparison with benchmark values \cite{ASP11}. The fine-structure constant is $\alpha=1/137.035\,999\,11$ and the atomic unit of time is $\hbar/E_h=2.418\,885 \times 10^{-17}$ s. $\Delta_{l-v}$ stands for the relative difference between length and velocity gauge values. Powers of ten are indicated within brackets.}
\begin{center}
\begin{tabular}{l l l l l l l}
\hline
\hline
           & \multicolumn{6}{c}{$2s_{1/2} \rightarrow 1s_{1/2}$ two-photon partial decay rates (s$^{-1}$)}                                                  \\
\hline
           & \multicolumn{2}{l}{Lagrange mesh}            & \multicolumn{2}{l}{$B$ polynomials \cite{ASP11}} & \multicolumn{2}{l}{$B$~splines \cite{ASP11}} \\
Multipoles \hspace{1cm} & Length gauge \hspace{1cm} & $\Delta_{l-v}$ \hspace{1cm} & Length gauge \hspace{1cm} & $\Delta_{l-v}$ \hspace{1cm} & Length gauge \hspace{1cm} & $\Delta_{l-v}$ \\
\hline
           & \multicolumn{6}{c}{$Z=1$}                                                                                                                      \\
$2E1$      & $8.229\,059\,158\,6$        & $<1.0\,[-15]$  & $8.229\,059\,158\,6$        & $<1.0\,[-26]$      & $8.229\,059\,150\,9$        & $<1.0\,[-15]$  \\
$E1M2$     & $2.537\,180\,773\,5\,[-10]$ & $<1.0\,[-15]$  & $2.537\,180\,773\,5\,[-10]$ & $<1.0\,[-25]$      & $2.537\,180\,763\,5\,[-10]$ & $<1.0\,[-15]$  \\
$2M1$      & $1.380\,358\,049\,6\,[-11]$ & \qquad $-$     & $1.380\,358\,049\,6\,[-11]$ & \qquad $-$         & $1.380\,358\,047\,3\,[-11]$ & \qquad $-$     \\
$2E2$      & $4.907\,228\,923\,3\,[-12]$ & $<1.0\,[-14]$  & $4.907\,228\,923\,2\,[-12]$ & $<1.0\,[-34]$      & $4.907\,228\,916\,5\,[-12]$ & $<1.0\,[-14]$  \\
$2M2$      & $3.069\,351\,007\,3\,[-22]$ & \qquad $-$     & $3.069\,351\,007\,4\,[-22]$ & \qquad $-$         & $3.069\,350\,983\,3\,[-22]$ & \qquad $-$     \\
$E2M1$     & $1.639\,7\,[-23]$           & $<1.0\,[-4]$   & $1.639\,356\,519\,7\,[-23]$ & $<1.0\,[-34]$      & $1.639\,741\,353\,0\,[-23]$ & $<1.0\,[-4]$   \\
Total      & $8.229\,059\,158\,9$        &                & $8.229\,059\,158\,9$        &                    & $8.229\,059\,151\,2$        &                \\
           & \multicolumn{6}{c}{$Z=40$}                                                                                                                     \\
$2E1$      & $3.198\,635\,6[+10]$        & $<1.0\,[-15]$  & $3.198\,62\,[+10]$          & $<1.0\,[-13]$      & $3.198\,58\,[+10]$          & $<1.0\,[-15]$  \\           
           & \multicolumn{6}{c}{$Z=92$}                                                                                                                     \\
$2E1$      & $3.825\,898\,[+12]$         & $<1.0\,[-12]$  & $3.825\,839\,[+12]$         & $<1.0\,[-9]$       & $3.825\,55\,[+12]$          & $<1.0\,[-15]$  \\
\hline
\hline
\end{tabular}
\end{center}
\label{table_2s_1s_Coulomb}
\end{table}
\twocolumngrid

\section{Numerical results}
\label{sec:res}

\subsection{Hydrogenic atoms}
\label{subsec:Coulomb}

\subsubsection{$2s_{1/2} \rightarrow 1s_{1/2}$ transition}

We first compute the $2s_{1/2} \rightarrow 1s_{1/2}$ two-photon decay rates for the Dirac-Coulomb problem, where $V(r)=-Z/r$ in atomic units. \tablename{~\ref{table_2s_1s_Coulomb}} presents values for $Z=1$, 40, and 92. The results obtained in the length gauge are compared with a benchmark calculation presented in \Ref{ASP11}, involving Bernstein-polynomial ($B$-polynomial) and $B$-spline finite-basis sets. The Lagrange-mesh and $B$-spline results are obtained in double precision, while the $B$-polynomial values are obtained in quadruple precision. For each $Z$ value, a very small number of Lagrange functions, i.e., $N_i=N_f=N=6$, is enough to accurately describe the initial $2s_{1/2}$ and final $1s_{1/2}$ states. The number of mesh points for the intermediate states is $N'=40$ for each value of $\kappa'$. As an example, the $2E1$ pair of multipoles involves $p_{1/2}$ ($\kappa'=1$) and $p_{3/2}$ ($\kappa'=-2$) intermediate states in $2s_{1/2} \xrightarrow{E1} p_{1/2} \xrightarrow{E1} 1s_{1/2}$ and $2s_{1/2} \xrightarrow{E1} p_{3/2} \xrightarrow{E1} 1s_{1/2}$. The Gauss-Legendre quadrature over $\omega_1$ is performed with $N_{\omega_1}=15$ mesh points, in order to be consistent with the choice made in \Ref{ASP11}. The Lagrange-mesh results are in excellent agreement with the $B$-polynomial values. For $Z=1$, all the figures displayed in the two columns of \tablename{~\ref{table_2s_1s_Coulomb}} are indeed identical, except for the $E2M1$ contribution. For this pair of multipoles, the invariance between length and velocity gauge values is poor ($\Delta_{l-v}<10^{-4}$, as for the $B$-spline results), while $\Delta_{l-v}<10^{-23}$ for the $B$-polynomial results. This numerical problem is most likely due to a cancellation of two values that are very close. However, this contribution is so small ($\sim 10^{-23}$ s$^{-1}$) that it does not affect the total decay rate. For all other results displayed in \tablename{~\ref{table_2s_1s_Coulomb}}, the test of gauge invariance is successfully performed: $\Delta_{l-v}$ ranges from $10^{-13}$ down to $10^{-16}$. When $Z$ increases, some differences occur between Lagrange-mesh and $B$-polynomial results, but $\Delta_{l-v}$ is lower with the Lagrange-mesh calculations.

In \tablename{~\ref{table_2s_1s_Coulomb}}, the use of a truncated value $2.418\,885 \times 10^{-17}$ s for the atomic unit of time, as in \Ref{ASP11}, leads to a difference on the sixth digit of the results. Besides, using 15 mesh points for the Gauss-Legendre quadrature over $\omega_1$ may be insufficient to reach the convergence of all the displayed figures. Hence, one has to investigate the stability of the figures with respect to a variation of the number of mesh points $N'$ and $N_{\omega_1}$.

\tablename{~\ref{table_2s_1s_Coulomb2}} displays the most significant multipole contributions (in s$^{-1}$) to the $2s_{1/2} \rightarrow 1s_{1/2}$ decay rate (in the velocity gauge) of hydrogenic ions with $Z=1$, 40, and 92. Both one-photon $M1$ and two-photon contributions to the total decay rate are presented. For $M1$ decay rates, $N=6$ mesh points are also used for the initial and final states, and the Gauss-Laguerre quadrature is performed with $N_G=20$ mesh points. For two-photon decay rates, the required number of mesh points $N'$ for the intermediate states increases with $Z$, from $N'=20$ ($Z=1$) to $N'=30$ ($Z=40$) and $N'=40$ ($Z=92$). Indeed, the difference $2(\gamma'-|\kappa'|)$ appearing as a power of $r$ in \Eq{Lag.3} increases more and more as $Z$ increases, since $\gamma'=\sqrt{\kappa'^2-(\alpha Z)^2}$. Hence, one needs higher $N'$ (and thus $N_G$) values to improve the accuracy of the Gauss-Laguerre quadrature in Eqs.~\eqref{Lag.10} to \eqref{Lag.12}. The required number of mesh points $N_{\omega_1}$ for the Gauss-Legendre quadrature over $\omega_1$ depends on the pair of multipoles, i.e., on the shape of the differential decay rate to be integrated over $\omega_1$.
\begin{table}[ht!]
\caption{{\footnotesize Multipole contributions (in s$^{-1}$) to the total $2s_{1/2} \rightarrow 1s_{1/2}$ decay rate of hydrogenic ions with $Z=1$, 40, and 92. `Total $2\gamma$' stands for the sum of the two-photon contributions, `Total' for the sum of `Total $2\gamma$' and the one-photon $M1$ decay rate. $\Delta_{l-v}$ stands for the relative difference between the length and velocity gauge values, $\Delta_{i-c}$ for the relative difference between incoherent and coherent computations. Powers of ten are indicated within brackets.}}
\begin{center}
\begin{tabular}{l l l l}
\hline
\hline
                & \multicolumn{3}{c}{$2s_{1/2} \rightarrow 1s_{1/2}$ partial decay rates (s$^{-1}$)} \\
\hline
Multipoles      & Velocity gauge                 & $\Delta_{l-v}$ & $\Delta_{i-c}$ \\
\hline
                & \multicolumn{3}{c}{$Z=1$}                                        \\
$2E1$           & $8.229\,061\,48$               & $<1.0\,[-15]$  & $<1.0\,[-13]$  \\
$E1M2$          & $2.537\,181\,50\,[-10]$        & $<1.0\,[-15]$  & $<1.0\,[-13]$  \\
$2M1$           & $1.380\,358\,437\,58\,[-11]$   & \qquad $-$     & \qquad $-$     \\
$2E2$           & $4.907\,230\,300\,47\,[-12]$   & $<1.0\,[-14]$  & $<1.0\,[-12]$  \\
$2M2$           & $3.069\,351\,872\,51\,[-22]$   & \qquad $-$     & \qquad $-$     \\
$E2M3$          & $1.422\,933\,545\,35\,[-22]$   & $<1.0\,[-15]$  & $<1.0\,[-11]$  \\
$E2M1$          & $1.639\,5\,[-23]$              & $<1.0\,[-4]$   & $<1.0\,[-4]$   \\
Total $2\gamma$ & $8.229\,061\,48$               &                &                \\
$M1$            & $2.495\,923\,647\,[-6]$        & \qquad $-$     & \qquad $-$     \\
Total           & $8.229\,063\,98$               &                &                \\
                & \multicolumn{3}{c}{$Z=40$}                                       \\
$2E1$           & $3.198\,672\,93\,[+10]$        & $<1.0\,[-15]$  & $<1.0\,[-8]$   \\
$E1M2$          & $2.532\,719\,173\,[+6]$        & $<1.0\,[-15]$  & $<1.0\,[-12]$  \\
$2M1$           & $1.566\,111\,378\,86\,[+5]$    & \qquad $-$     & \qquad $-$     \\
$2E2$           & $5.004\,194\,969\,[+4]$        & $<1.0\,[-14]$  & $<1.0\,[-11]$  \\
$2M2$           & $8.226\,285\,305\,5$           & \qquad $-$     & \qquad $-$     \\
$E2M3$          & $3.888\,921\,643\,8$           & $<1.0\,[-15]$  & $<1.0\,[-11]$  \\
$E2M1$          & $5.491\,783\,844\,[-1]$        & $<1.0\,[-10]$  & $<1.0\,[-10]$  \\
Total $2\gamma$ & $3.198\,946\,87\,[+10]$        &                &                \\
$M1$            & $2.874\,694\,006\,737\,[+10]$  & \qquad $-$     & \qquad $-$     \\
Total           & $6.073\,640\,88\,[+10]$        &                &                \\
                & \multicolumn{3}{c}{$Z=92$}                                       \\
$2E1$           & $3.825\,894\,[+12]$            & $<1.0\,[-13]$  & $<1.0\,[-6]$   \\
$E1M2$          & $9.173\,467\,780\,9\,[+9]$     & $<1.0\,[-15]$  & $<1.0\,[-12]$  \\
$2M1$           & $1.118\,272\,687\,14\,[+9]$    & \qquad $-$     & \qquad $-$     \\
$2E2$           & $1.793\,657\,299\,[+8]$        & $<1.0\,[-14]$  & $<1.0\,[-12]$  \\
$2M2$           & $9.971\,596\,258\,4\,[+5]$     & \qquad $-$     & \qquad $-$     \\
$E2M3$          & $4.987\,722\,592\,9\,[+5]$     & $<1.0\,[-15]$  & $<1.0\,[-13]$  \\
$E2M1$          & $1.789\,081\,429\,6\,[+5]$     & $<1.0\,[-12]$  & $<1.0\,[-10]$  \\
Total $2\gamma$ & $3.836\,366\,[+12]$            &                &                \\
$M1$            & $1.946\,812\,787\,741\,3[+14]$ & \qquad $-$     & \qquad $-$     \\
Total           & $1.985\,176\,45\,[+14]$        &                &                \\
\hline
\hline
\end{tabular}
\end{center}
\label{table_2s_1s_Coulomb2}
\end{table}
The values range from $N_{\omega_1}=20$ ($2M1$) to $N_{\omega_1}=40$ or 60 ($2E1$, $E1M2$). The number of significant figures then depends on both the nuclear charge $Z$ and the pair of multipoles. It ranges from seven ($2E1$ with $Z=92$) up to twelve for the most accurate results. The same exception as in \tablename{~\ref{table_2s_1s_Coulomb}} occurs, namely, only five figures are stable for $E2M1$ with $Z=1$. However, as $Z$ increases its number of significant figures increases and $\Delta_{l-v}$ decreases, down to $10^{-13}$ for $Z=92$. For all the other results an excellent $\Delta_{l-v}$ is found, from $10^{-14}$ to $10^{-16}$.

Another test of the accuracy is given by the value of $\Delta_{i-c}$, which stands for the relative difference between incoherent and coherent calculations in the length gauge. As explained in Sec.~\ref{sec:formulation}, the former implies a sum of the squares of reduced matrix elements for electric and longitudinal types of transition according to $[S_{1,1}^{j_\nu}]^2+[S_{1,-1}^{j_\nu}]^2+[ S_{-1,1}^{j_\nu}]^2+[S_{-1,-1}^{j_\nu}]^2$, while the latter implies a squared sum of reduced matrix elements for electric type of transition according to $[S_{1,1}^{j_\nu}+S_{1,-1}^{j_\nu}+S_{-1,1}^{j_\nu}+S_{-1,-1}^{j_\nu}]^2$. The coherent summation provides information about the accuracy of the numerical results. According to the general requirement of gauge invariance, an exact calculation of $S_{1,-1}^{j_\nu}$, $S_{-1,1}^{j_\nu}$ and $S_{-1,-1}^{j_\nu}$ should give identically zero, and both incoherent and coherent summations should therefore give the same results. Let us denote by $\epsilon$ the numerical error made on the computation of these three terms. For an incoherent summation, the error becomes of the order of $\epsilon^2$ for the differential decay rate, because the sum of the squares is $[S_{1,1}^{j_\nu}]^2+O(\epsilon^2)$. The error is much smaller than the initial error $\epsilon$. By contrast, for a coherent summation, a supplementary error of the order of $\epsilon$ is induced by the cross terms appearing in the square of the sum, because the latter is $[S_{1,1}^{j_\nu}+O(\epsilon)]^2=[S_{1,1}^{j_\nu}]^2+O(\epsilon)$. The total error is then larger and $\Delta_{i-c}$ provides a lower bound for the actual numerical error. The value of $\Delta_{i-c}$ in \tablename{~\ref{table_2s_1s_Coulomb2}} is in any case higher than $\Delta_{l-v}$, but does not affect the number of significant digits determined by variations of $N'$ and $N_{\omega_1}$.

For $Z=1$, the two-photon contribution to the total decay rate is dominant, but as $Z$ increases the competition between both processes is evolving towards a domination of the one-photon $M1$ contribution, as shown in \figurename{~\ref{fig_2gamma_M1_Coulomb}}. For $Z=40$, both processes show decay rates of the same order of magnitude. The competition inside two-photon processes is also evolving with $Z$, as already investigated in \Ref{SPI98}. The same conclusions about the total decay rates are found with the present Lagrange-mesh calculation, which provides accuracies improved by two orders of magnitude.

\tablename{~\ref{table_2s_1s_Coulomb3}} displays the $2s_{1/2} \rightarrow 1s_{1/2}$ total decay rates (in s$^{-1}$) of hydrogenic ions with selected $Z$ values up to 100. Enough multipoles are included in the calculation of the total two-photon decay rates to reach an accuracy of at least seven figures. For the total decay rates, an accuracy of nine figures is found for all displayed values.
\begin{table}[ht!]
\caption{{\footnotesize Total $2s_{1/2} \rightarrow 1s_{1/2}$ decay rates (in s$^{-1}$) of hydrogenic ions with given $Z$ values from 1 to 100. `Total $2\gamma$' stands for the sum of the two-photon contributions, `Total' for the sum of `Total $2\gamma$' and the one-photon $M1$ decay rate. Powers of ten are indicated within brackets.}}
\begin{center}
\resizebox{0.485\textwidth}{!}{
\begin{tabular}{l l l l}
\hline
\hline
    & \multicolumn{3}{c}{$2s_{1/2} \rightarrow 1s_{1/2}$ decay rates (s$^{-1}$)}        \\
\hline
$Z$ & Total $2\gamma$         & $M1$                          & Total                   \\
\hline
1   & $8.229\,061\,48$        & $2.495\,923\,647\,[-6]$       & $8.229\,063\,98$        \\
2   & $5.266\,041\,42\,[+2]$  & $2.556\,261\,874\,[-3]$       & $5.266\,066\,98\,[+2]$  \\
5   & $1.284\,704\,54\,[+5]$  & $2.440\,755\,638\,[+1]$       & $1.284\,948\,62\,[+5]$  \\
10  & $8.200\,643\,91\,[+6]$  & $2.510\,040\,967\,5\,[+4]$    & $8.225\,744\,32\,[+6]$  \\
20  & $5.195\,165\,68\,[+8]$  & $2.614\,937\,321\,472\,[+7]$  & $5.456\,659\,41\,[+8]$  \\
30  & $5.821\,288\,66\,[+9]$  & $1.552\,527\,288\,200\,[+9]$  & $7.373\,815\,94\,[+9]$  \\
40  & $3.198\,946\,87\,[+10]$ & $2.874\,694\,006\,737\,[+10]$ & $6.073\,640\,88\,[+10]$ \\
50  & $1.186\,912\,90\,[+11]$ & $2.830\,330\,252\,000\,[+11]$ & $4.017\,243\,16\,[+11]$ \\
60  & $3.428\,240\,8\,[+11]$  & $1.881\,362\,957\,514\,[+12]$ & $2.224\,187\,03\,[+12]$ \\
70  & $8.314\,130\,0\,[+11]$  & $9.603\,749\,351\,584\,[+12]$ & $1.043\,516\,23\,[+13]$ \\
80  & $1.770\,223\,3\,[+12]$  & $4.073\,704\,104\,592\,[+13]$ & $4.250\,726\,43\,[+13]$ \\
90  & $3.402\,365\,[+12]$     & $1.515\,142\,818\,980\,[+14]$ & $1.549\,166\,46\,[+14]$ \\
100 & $6.005\,549\,[+12]$     & $5.149\,350\,767\,050\,[+14]$ & $5.209\,406\,26\,[+14]$ \\
\hline
\hline
\end{tabular}
}
\end{center}
\label{table_2s_1s_Coulomb3}
\end{table}
\begin{figure}[ht!]
\begin{center}
\includegraphics[scale=0.465]{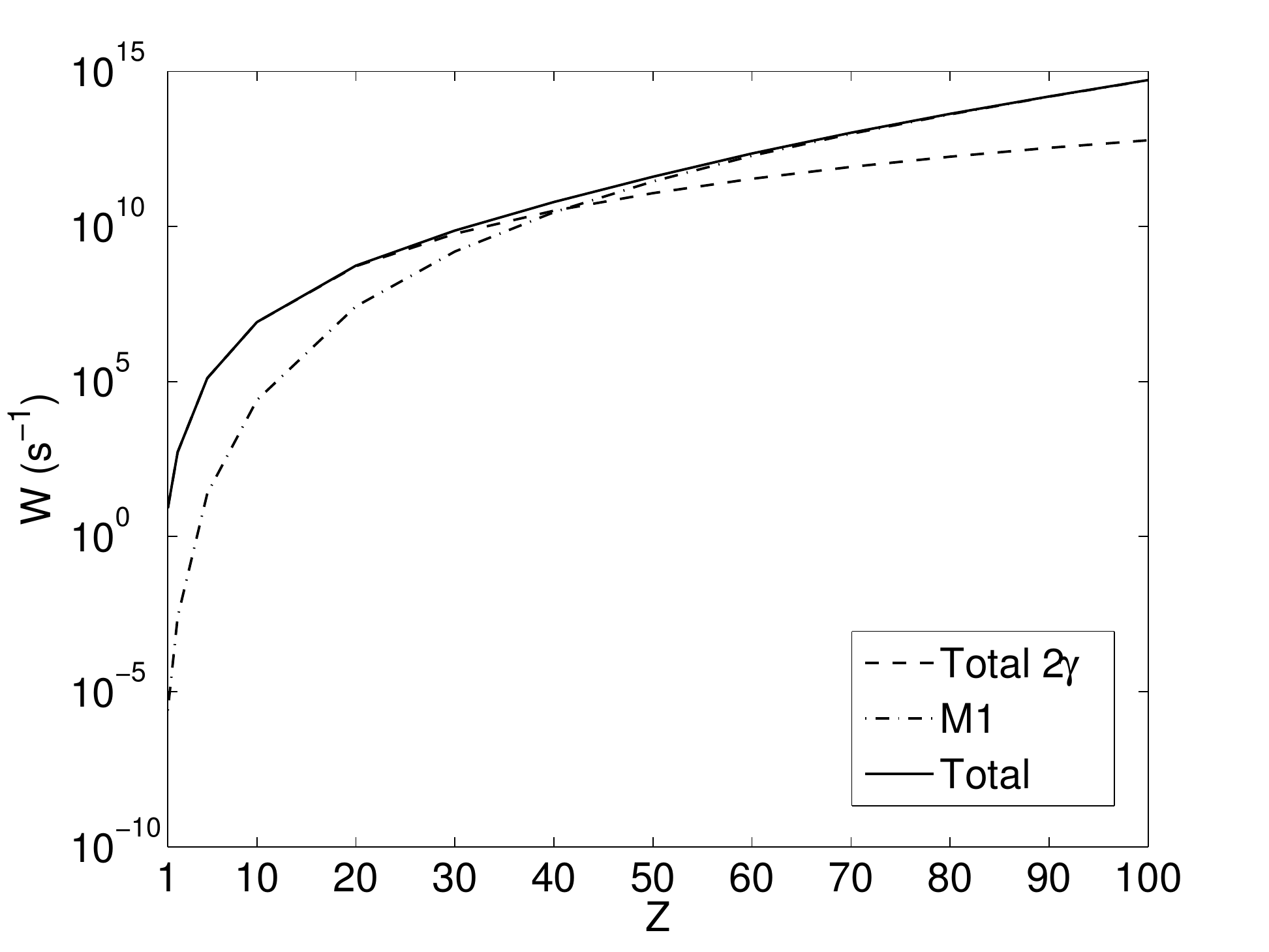}
\end{center}
\caption{{\footnotesize $2s_{1/2} \rightarrow 1s_{1/2}$ decay rate (in s$^{-1}$) as a function of $Z$. `Total $2\gamma$' (dashed line) stands for the sum of the two-photon contributions, `Total' (solid line) for the sum of `Total $2\gamma$' and $M1$ (dashed-dotted line).}}
\label{fig_2gamma_M1_Coulomb}
\end{figure}

%%%%%%%%%%%%%%%%%%%%%%%%%%%%%%%%%%%%%%%%%%%%%%%%%%%%%%%%%%%%%%%%%%%%%%%%%%%%%%%%%%%%%%%%%%%%%%%%%%%%%%%%%%%%%%%%%%%%%%%%%%

\subsubsection{$2p_{1/2} \rightarrow 1s_{1/2}$ transition}

We then compute the $2p_{1/2} \rightarrow 1s_{1/2}$ two-photon decay rates for hydrogenic ions. As an example of intermediate states appearing in this transition, the $E1E2$ pair of multipoles involves $d_{3/2}$ ($\kappa'=2$) states in $2p_{1/2} \xrightarrow{E1} d_{3/2} \xrightarrow{E2} 1s_{1/2}$, while $E2E1$ involves $p_{3/2}$ ($\kappa'=-2$) states in $2p_{1/2} \xrightarrow{E2} p_{3/2} \xrightarrow{E1} 1s_{1/2}$. Although these two pairs of multipoles do not involve the same intermediate states, their decay rates integrated over $\omega_1$ give the same value.

\tablename{~\ref{table_2p_1s_Coulomb}} displays the most significant multipole contributions (in s$^{-1}$) to the $2p_{1/2} \rightarrow 1s_{1/2}$ two-photon decay rate (in the velocity gauge) of hydrogenic ions with $Z=1$, $40$, and $92$. Both one-photon $E1$ and two-photon contributions to the total decay rate are presented. Enough multipoles are included in the calculation of the total two-photon decay rates to reach an accuracy of ten figures. The two-photon results are compared with benchmark calculations involving Dirac-Coulomb Sturmians~\cite{BTE14} and $B$-spline finite-basis sets~\cite{ASP09}. $E2M2$ and $E2E3$ pairs of multipoles are not considered in \Ref{BTE14}.

The same value of $N$ as in \tablename{~\ref{table_2s_1s_Coulomb}} is used, i.e., $N=6$. With $N_{\omega_1}=40$ for the Gauss-Legendre quadrature, the Lagrange-mesh results are in excellent agreement with the Sturmian values from \Ref{BTE14}. For each $Z$ value, all the figures displayed in the two columns of \tablename{~\ref{table_2p_1s_Coulomb}} are identical. \Ref{ASP09} uses $N_{\omega_1}=15$ mesh points and the truncated value $2.418\,885 \times 10^{-17}$~s for the atomic unit of time. With these choices, the Lagrange-mesh results reproduce the $B$-spline values for $Z=1$, but differ more and more as $Z$ increases. The difference appears on the fourth digit for $Z=40$, and on the third digit for $Z=92$. For all the Lagrange-mesh results displayed in \tablename{~\ref{table_2p_1s_Coulomb}}, the test of gauge invariance gives $\Delta_{l-v}<10^{-14}$, to be compared with $\Delta_{l-v}<10^{-10}$ in \Ref{BTE14}.

As in \tablename{~\ref{table_2s_1s_Coulomb2}}, one has to investigate the stability of the figures for the $2p_{1/2} \rightarrow 1s_{1/2}$ transition with respect to a variation of $N'$ and $N_{\omega_1}$. The number $N'$ increases from 20 ($Z=1$) to 30 ($Z=40$) and 40 ($Z=92$), while $N_{\omega_1}$ varies from 40 to 60 depending on the pairs of multipoles considered. The number of significant figures ranges from nine ($M1M2$ with $Z=1$) up to eleven for the most accurate results. As in \tablename{~\ref{table_2s_1s_Coulomb2}}, $\Delta_{i-c}$ is higher than $\Delta_{l-v}$, but does not affect the number of significant digits.

In comparison with two-photon processes, the order of magnitude of the one-photon $E1$ decay rate is from five ($Z=92$) to thirteen ($Z=1$) times higher, which illustrates its dominance in the decay of the $2p_{1/2}$ state.

%%%%%%%%%%%%%%%%%%%%%%%%%%%%%%%%%%%%%%%%%%%%%%%%%%%%%%%%%%%%%%%%%%%%%%%%%%%%%%%%%%%%%%%%%%%%%%%%%%%%%%%%%%%%%%%%%%%%%%%%%%%%%%%%%%%%%%%%%%%%%%%%%%%%%%%%%%%%%%%%%%%%%%%%%%%%%%%%%%%%%%%%%%%%%%%%%%%%%%%%%%%%%%%%%%%%%%%%%%%%%%%%%%%%%%%%%%%%%%%%%%%%

\subsection{Yukawa potential}
\label{subsec:Yukawa}

Two-photon decay rates can also be accurately computed for Yukawa potentials 
\beq
V(r) = -V_0\, \frac{e^{-\mu r}}{r},
\eeqn{Yuk:eq1}
with different values of $V_0$ and $\mu$. Within the Lagrange-mesh method, switching to Yukawa potentials only requires to change the potential values $V(hx_i)$ in the Hamiltonian matrix given by \Eq{Lag.7}. Also for this kind of potentials, it is shown in \Ref{BFG14} that the Lagrange-mesh method is able to provide accurate energies with a number of mesh points for which the computation seems instantaneous. The approximate wave functions provide mean values of powers of the coordinate that are also extremely precise. \Ref{FGB14} shows that accurate static dipole polarizabilities can be obtained for Yukawa potentials with the Lagrange-mesh method, for the ground state as well as for excited states.
\clearpage
\onecolumngrid

\begin{table}[ht!]
\caption{\footnotesize Multipole contributions (in $s^{-1}$) to the $2p_{1/2} \rightarrow 1s_{1/2}$ two-photon decay rate of hydrogenic ions with $Z=1$, $40$, and $92$. Comparison with benchmark values~\cite{BTE14} and~\cite{ASP09}. `Total $2\gamma$' stands for the sum of the two-photon contributions, `Total' for the sum of `Total $2\gamma$' and the one-photon $E1$ decay rate. $\Delta_{l-v}$ stands for the relative difference between the length and velocity gauge values, $\Delta_{i-c}$ for the relative difference between incoherent and coherent computations. Powers of ten are indicated within brackets.}
\begin{center}
\begin{tabular}{l l l l l l}
\hline
\hline
                & \multicolumn{5}{c}{$2p_{1/2} \rightarrow 1s_{1/2}$ partial decay rates (s$^{-1}$)}                                     \\
\hline
                & \multicolumn{3}{c}{Lagrange-mesh} \hspace{2.5cm}                   & Sturmians~\cite{BTE14} & $B$~splines~\cite{ASP09} \\
Multipoles \hspace{1cm} & Velocity gauge \hspace{1cm} & $\Delta_{l-v}$ \hspace{1cm} & $\Delta_{i-c}$ \hspace{2.5cm} & Velocity gauge \hspace{1.5cm} & Velocity gauge \\
\hline
                & \multicolumn{5}{c}{$Z=1$}                                                                                              \\
$E1M1$          & $9.676\,656\,889\,[-6]$          & $<1.0\,[-15]$  & $<1.0\,[-13]$  & $9.676\,656\,[-6]$     & $9.676\,654\,[-6]$       \\
$E1E2$          & $6.611\,798\,085\,82\,[-6]$      & $<1.0\,[-15]$  & $<1.0\,[-13]$  & $6.611\,798\,[-6]$     & $6.611\,79\,[-6]$        \\
$E2M2$          & $9.385\,472\,823\,37\,[-17]$     & $<1.0\,[-15]$  & $<1.0\,[-12]$  &                        & $9.385\,470\,[-17]$      \\
$M1M2$          & $3.827\,878\,99\,[-17]$          & \qquad $-$     & \qquad $-$     & $3.827\,879\,[-17]$    & $3.827\,877\,[-17]$      \\
$E2E3$          & $4.095\,986\,879\,01\,[-18]$     & $<1.0\,[-14]$  & $<1.0\,[-12]$  &                        & $4.095\,985\,[-18]$      \\
Total $2\gamma$ & $1.628\,845\,497\,[-5]$          &                &                & $1.628\,845\,[-5]$     & $1.628\,844\,[-5]$       \\
$E1$            & $6.268\,354\,359\,740\,7\,[+8]$  & $<1.0\,[-14]$  & \qquad $-$     &                        &                          \\
Total           & $6.268\,354\,359\,740\,9\,[+8]$  &                &                &                        &                          \\            
                & \multicolumn{5}{c}{$Z=40$}                                                                                             \\
$E1M1$          & $6.027\,545\,915\,[+7]$          & $<1.0\,[-15]$  & $<1.0\,[-11]$  & $6.027\,546\,[+7]$     & $6.027\,323\,[+7]$       \\
$E1E2$          & $4.092\,510\,063\,39\,[+7]$      & $<1.0\,[-15]$  & $<1.0\,[-11]$  & $4.092\,510\,[+7]$     & $4.092\,020\,[+7]$       \\
$E2M2$          & $1.521\,964\,162\,55\,[+3]$      & $<1.0\,[-15]$  & $<1.0\,[-12]$  &                        & $1.521\,687\,[+3]$       \\
$M1M2$          & $5.603\,699\,503\,[+2]$          & \qquad $-$     & \qquad $-$     & $5.603\,699\,[+2]$     & $5.602\,320\,[+2]$       \\
$E2E3$          & $6.608\,670\,915\,8\,[+1]$       & $<1.0\,[-14]$  & $<1.0\,[-11]$  &                        & $6.608\,612\,[+1]$       \\
Total $2\gamma$ & $1.012\,027\,082\,[+8]$          &                &                & $1.012\,011\,[+8]$     & $1.011\,940\,[+8]$       \\
$E1$            & $1.620\,964\,973\,742\,0\,[+15]$ & $<1.0\,[-14]$  & \qquad $-$     &                        &                          \\
Total           & $1.620\,965\,074\,944\,7\,[+15]$ &                &                &                        &                          \\
                & \multicolumn{5}{c}{$Z=92$}                                                                                             \\
$E1M1$          & $3.876\,927\,423\,[+10]$         & $<1.0\,[-15]$  & $<1.0\,[-10]$  & $3.876\,927\,[+10]$    & $3.863\,302\,[+10]$      \\
$E1E2$          & $2.374\,810\,767\,[+10]$         & $<1.0\,[-15]$  & $<1.0\,[-9]$   & $2.374\,811\,[+10]$    & $2.358\,404\,[+10]$      \\
$E2M2$          & $2.865\,365\,628\,3\,[+7]$       & $<1.0\,[-15]$  & $<1.0\,[-12]$  &                        & $2.834\,065\,[+7]$       \\
$M1M2$          & $7.753\,754\,779\,1\,[+6]$       & \qquad $-$     & \qquad $-$     & $7.753\,755\,[+6]$     & $7.689\,142\,[+6]$       \\
$E2E3$          & $1.178\,204\,267\,1\,[+6]$       & $<1.0\,[-14]$  & $<1.0\,[-12]$  &                        & $1.177\,403\,[+6]$       \\
Total $2\gamma$ & $6.255\,496\,752\,[+10]$         &                &                & $6.252\,513\,[+10]$    & $6.222\,474\,[+10]$      \\
$E1$            & $4.726\,013\,372\,756\,7\,[+16]$ & $<1.0\,[-15]$  & \qquad $-$     &                        &                          \\
Total           & $4.726\,019\,628\,253\,5\,[+16]$ &                &                &                        &                          \\
\hline
\hline
\end{tabular}
\end{center}
\label{table_2p_1s_Coulomb}
\end{table}
\twocolumngrid

Potential \rref{Yuk:eq1} has the singular behavior
\beq
V(r) \arrow{r}{0} - \frac{V_0}{r}
\eeqn{Yuk:eq2}
at the origin. 
Parameter $\gamma$ is thus given by \Eq{r.11} and parameter $\alpha$ is the same as in the Coulomb case, i.e., $\alpha=2(\gamma-\vert\kappa\vert)$. The scaling parameters $h_i,h_f,h'$ and the numbers of mesh points $N,N'$ are adjusted for each potential. Here we choose to use $h'=h_f$ as for the Coulomb case, and $N=N'$. Then $N_G=N=N'$ for the Gauss-Laguerre quadrature.

\subsubsection{$2s_{1/2} \rightarrow 1s_{1/2}$ transition}

\tablename{~\ref{table_2s_1s_M1_2E1_Yukawa}} lists the one-photon $M1$ and two-photon $2E1$ $2s_{1/2} \rightarrow 1s_{1/2}$ decay rates of a hydrogen atom embedded in a Debye plasma. Various values of the Debye length $\delta$ are considered. This situation is described by Yukawa potentials with $V_0=Z=1$ and $\mu=1/\delta$. Since $V_0=Z=1$, the $2E1$ contribution accurately gives the total two-photon decay rate. The limit $\delta \rightarrow \infty$ corresponds to the Coulomb case. For these $2E1$ decay rates, all computations are performed in the velocity gauge with given $N=N'$ mesh points, and the significant digits of the results are estimated by a comparison with $N+10$ mesh points. $N_{\omega_1}=80$ mesh points for the Gauss-Legendre quadrature are used to ensure at least eight significant figures for all $\mu$ values. The scaling parameter $h_{2s_{1/2}}$ starts from the Coulomb optimal value 1.0 and progressively increases with $\mu$, while $h_{1s_{1/2}}$ keeps the Coulomb optimal value 0.5 for all $\mu$ values. For $M1$ decay rates, $N_G=N$ mesh points are used for the Gauss-Laguerre quadrature, with the same $h_{2s_{1/2}}$ and $h_{1s_{1/2}}$ values. With this set of parameters, \: an \: excellent \: gauge \: invariance
\clearpage
\onecolumngrid

\begin{table}[ht!]
\caption{{\footnotesize $M1$ and $2E1$ $2s_{1/2} \rightarrow 1s_{1/2}$ decay rates (in s$^{-1}$) in the velocity gauge for Yukawa potentials with $V_0=Z=1$ and screening lengths $\delta=1/\mu$ (in a.u.). $E_{2s_{1/2}}$ and $E_{1s_{1/2}}$ are the energies (in a.u.) of the $2s_{1/2}$ and $1s_{1/2}$ states. Powers of ten are indicated within brackets.}}
\begin{center}
\begin{tabular}{l l l l l l l}
\hline
\hline
$\delta$ \hspace{0.375cm} & $N$ \hspace{0.375cm} & $h_{2s_{1/2}}$ \hspace{0.5cm} & $E_{2s_{1/2}}$ (a.u.) \hspace{1.75cm} & $E_{1s_{1/2}}$ (a.u.) \hspace{1.75cm} & $M1$ decay rate (s$^{-1}$) \hspace{0.5cm} & $2E1$ decay rate (s$^{-1}$) \\
\hline
$\infty$ & 40  & 1.0 & $-0.125\,002\,080\,189\,19$      & $-0.500\,006\,656\,596\,56$ & $2.495\,923\,6\,[-6]$ & $8.229\,061\,5$       \\
40       &     &     & $-0.101\,777\,950\,362\,96$      & $-0.475\,467\,842\,439\,41$ & $2.436\,008\,2\,[-6]$ & $8.024\,064\,8$       \\
20       &     &     & $-8.177\,315\,565\,392\,7\,[-2]$ & $-0.451\,823\,053\,642\,44$ & $2.277\,996\,1\,[-6]$ & $7.493\,121\,9$       \\
10       &     &     & $-4.992\,994\,282\,725\,0\,[-2]$ & $-0.407\,064\,567\,620\,90$ & $1.782\,761\,2\,[-6]$ & $5.881\,835\,1$       \\
5        &     &     & $-1.210\,873\,564\,573\,2\,[-2]$ & $-0.326\,814\,732\,823\,73$ & $6.796\,862\,2\,[-7]$ & $2.386\,905\,7$       \\
4        & 50  & 1.2 & $-3.396\,358\,802\,371\,[-3]$    & $-0.290\,925\,593\,773\,43$ & $2.818\,105\,9\,[-7]$ & $1.067\,773\,4$       \\
3.9      &     & 1.3 & $-2.689\,939\,159\,145\,[-3]$    & $-0.286\,516\,046\,440\,10$ & $2.416\,485\,5\,[-7]$ & $9.272\,593\,6\,[-1]$ \\
3.7      &     & 1.5 & $-1.456\,790\,715\,711\,[-3]$    & $-0.277\,126\,912\,819\,00$ & $1.633\,373\,0\,[-7]$ & $6.458\,338\,5\,[-1]$ \\
3.5      & 70  & 2.2 & $-5.373\,285\,510\,481\,[-4]$    & $-0.266\,895\,374\,665\,69$ & $8.953\,384\,[-8]$    & $3.677\,164\,8\,[-1]$ \\
3.4      & 80  & 2.5 & $-2.294\,292\,365\,537\,[-4]$    & $-0.261\,428\,990\,228\,90$ & $5.513\,432\,[-8]$    & $2.316\,753\,7\,[-1]$ \\
3.3      & 100 & 4.0 & $-4.516\,498\,773\,37\,[-5]$     & $-0.255\,708\,043\,484\,65$ & $2.290\,531\,[-8]$    & $9.879\,369\,8\,[-2]$ \\
\hline
\hline
\end{tabular}
\end{center}
\label{table_2s_1s_M1_2E1_Yukawa}
\end{table}
\twocolumngrid

\noindent $\Delta_{l-v}<10^{-15}$ is found for $2E1$ decay rates with all screening lengths. The number of mesh points $N=N'$ is increasing as $\delta$ decreases. Indeed, more mesh points are required in order to keep an excellent accuracy, as the upper $2s_{1/2}$ state of the transition becomes less and less bound when $\delta$ decreases, which affects the numerical results. One observes a decrease of both $M1$ and $2E1$ decay rates as the screening length $\delta$ decreases, down to two orders of magnitude lower for $\delta=3.3$ a.u. than for the Coulomb case. The plasma density scales as $1/\delta^3$, implying that the denser the plasma is, the lower the two-photon decay rates will be. For $\delta<3.3$ a.u. the $2s_{1/2}$ state is not bound any more. The ratio between the $M1$ and $2E1$ decay rates decreases when $\mu$ increases, but keeps the same order of magnitude: $W(M1)/W(2E1) \sim 10^{-7}$.

\figurename{~\ref{fig_2E1_Yukawa}} shows the spectral distribution function $\psi(y,V_0=1)$ of the $2E1$ $2s_{1/2} \rightarrow 1s_{1/2}$ transition for the Coulomb potential (solid line) and four Yukawa potentials with given $\mu$ values (in a.u.).
\begin{figure}[ht!]
\begin{center}
\includegraphics[scale=0.4725]{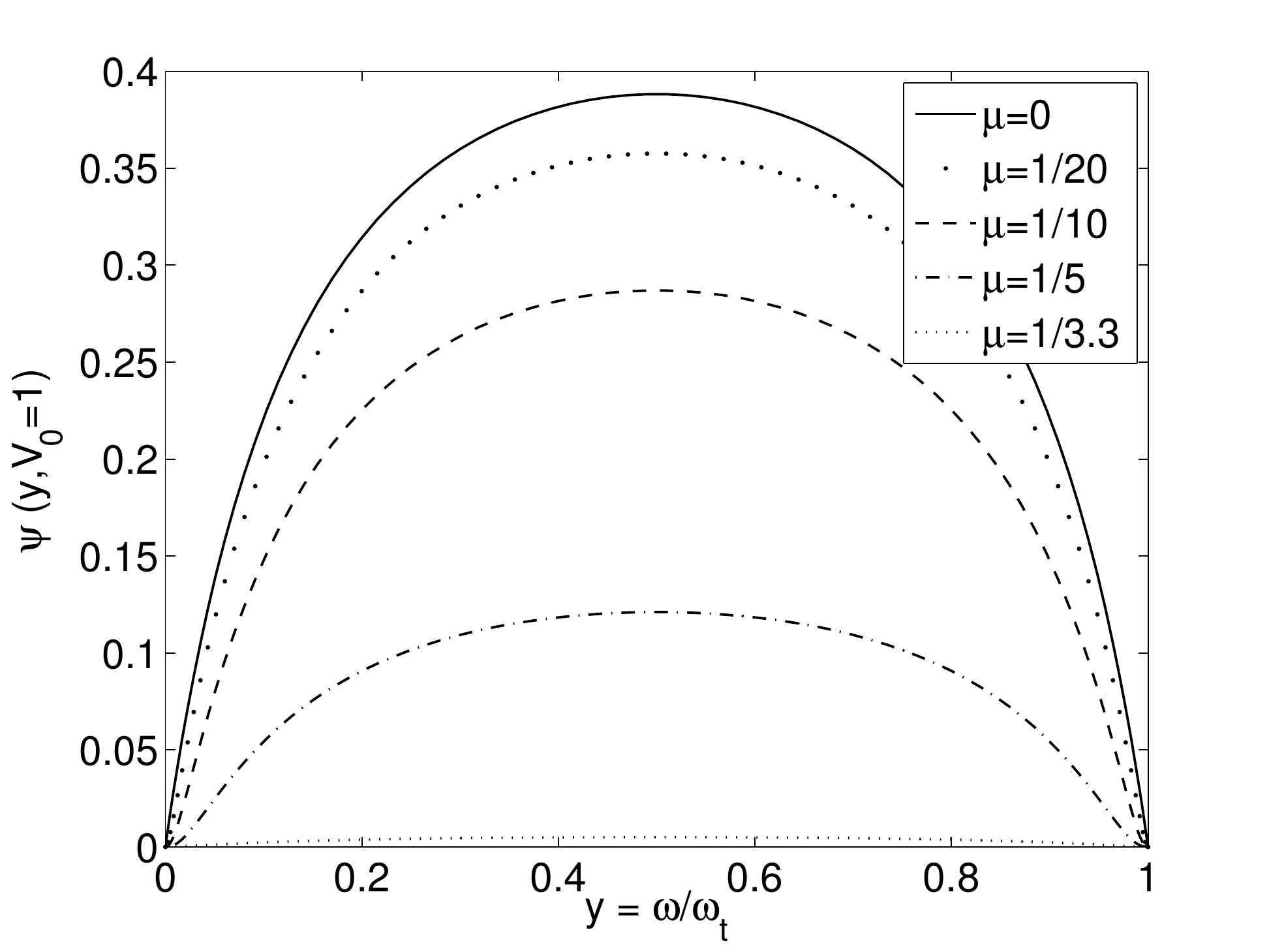}
\end{center}
\caption{{\footnotesize Spectral distribution function $\psi(y,V_0=1)$ of the $2E1$ $2s_{1/2} \rightarrow 1s_{1/2}$ transition for Yukawa potentials with $V_0=1$ and given $\mu$ values (in a.u.). The variable $y=\omega/\omega_t$ is the fraction of the photon energy carried by the first photon.}}
\label{fig_2E1_Yukawa}
\end{figure}
For a given pair of multipoles $\lambda_1 L_1,\lambda_2 L_2$ the function $\psi(y,V_0)$ is defined as~\cite{GD81}
\beq
\dfrac{d\overline{W}_{\lambda_1 L_1,\lambda_2 L_2}}{dy} = \left( \frac{9}{2^{10}} \right) (V_0/c)^n \psi(y,V_0),
\eeqn{eq_spectral_distr}
where $y=\omega/\omega_t$ is the fraction of the photon energy carried by one of the photons and $\omega_t$ is the energy of the transition. Both $\psi(y,V_0)$ and $y$ are dimensionless quantities. For the $2E1$ contribution, parameter $n$ equals 6 in \Eq{eq_spectral_distr}. As $2E1$ involves two photons with the same characteristics, one expects $\psi(y,V_0=1)$ to be symmetric around $y=0.5$, as shown in \figurename{~\ref{fig_2E1_Yukawa}}. From the Coulomb case $\mu=0$ to the case $\mu=1/3.3$ a.u., the maximum $\psi(0.5,V_0=1)$ decreases from 0.4 to 0.005.

%%%%%%%%%%%%%%%%%%%%%%%%%%%%%%%%%%%%%%%%%%%%%%%%%%%%%%%%%%%%%%%%%%%%%%%%%%%%%%%%%%%%%%%%%%%%%%%%%%%%%%%%%%%%%%%%%%%%%%%%%%%

\subsubsection{$2p_{1/2} \rightarrow 1s_{1/2}$ transition}

The influence of the screening length $\delta$ on the $E1M1$ and $E1E2$ $2p_{1/2} \rightarrow 1s_{1/2}$ decay rates of a hydrogen atom embedded in a Debye plasma is studied in \tablename{~\ref{table_2p_1s_E1M1_E1E2_Yukawa}}. All computations are performed in the velocity gauge with given $N=N'$ mesh points, and the significant digits of the results are estimated by a comparison with $N+10$ mesh points. As for the $2s_{1/2} \rightarrow 1s_{1/2}$ transition, $N_{\omega_1}=80$ is chosen to ensure at least nine significant figures for all $\mu$ values. The scaling parameter $h_{2p_{1/2}}$ starts from 1.0 and progressively increases with $\mu$, while $h_{1s_{1/2}}$ is 0.5 for all $\mu$ values. With this set of parameters, \: an \: excellent
\clearpage
\onecolumngrid

\begin{table}[ht!]
\caption{{\footnotesize Non-resonant $E1M1^\ast$ and $E1E2$ $2p_{1/2} \rightarrow 1s_{1/2}$ decay rates (in s$^{-1}$) in the velocity gauge for Yukawa potentials with $V_0=1$ and screening lengths $\delta=1/\mu$ (in a.u.). $E_{2p_{1/2}}$ and $E_{1s_{1/2}}$ are the energies (in a.u.) of the $2p_{1/2}$ and $1s_{1/2}$ states, and $y_1^R=\omega_1^R/\omega_t$, where $\omega_1^R=E_{2p_{1/2}}-E_{2s_{1/2}}$ and $\omega_t=E_{2p_{1/2}}-E_{1s_{1/2}}$, is the fraction of the photon energy carried by the first photon at the resonance associated with the $2s_{1/2}$ state. Powers of ten are indicated within brackets.}}
\begin{center}
\begin{tabular}{l l l l l l l l}
\hline
\hline
$\delta$ \hspace{0.25cm} & $N$ \hspace{0.25cm} & $h_{2p_{1/2}}$ & $E_{2p_{1/2}}$ (a.u.) \hspace{0.75cm} & $E_{1s_{1/2}}$ (a.u.) \hspace{1.25cm} & $y_1^R$ \hspace{0.75cm} & $E1M1^\ast$ decay rate (s$^{-1}$) & $E1E2$ decay rate (s$^{-1}$) \\
\hline
$\infty$ & 40  & 1.0            & $-0.125\,002\,080\,189\,19$      & $-0.500\,006\,656\,596\,56$  & \quad $-$   & $9.676\,656\,89\,[-6]$ & $6.611\,798\,086\,[-6]$ \\
40       &     &                & $-0.101\,494\,510\,284\,31$      & $-0.475\,467\,842\,439\,41$  & $7.6\,[-4]$ & $9.482\,347\,81\,[-6]$ & $6.539\,687\,823\,[-6]$ \\
20       &     &                & $-8.074\,234\,289\,207\,9\,[-2]$ & $-0.451\,823\,053\,642\,44$  & $2.8\,[-3]$ & $8.957\,461\,75\,[-6]$ & $6.336\,966\,360\,[-6]$ \\
10       &     &                & $-4.653\,603\,444\,379\,0\,[-2]$ & $-0.407\,064\,567\,620\,90$  & $9.4\,[-3]$ & $7.210\,479\,53\,[-6]$ & $5.595\,257\,319\,[-6]$ \\
5        &     &                & $-4.102\,311\,012\,023\,[-3]$    & $-0.326\,814\,732\,823\,73$  & $2.5\,[-2]$ & $2.465\,333\,47\,[-6]$ & $2.793\,692\,534\,[-6]$ \\
4.9      & 50  & 1.2            & $-3.114\,540\,304\,660\,[-3]$    & $-0.323\,783\,677\,162\,65$  & $2.5\,[-2]$ & $2.251\,154\,97\,[-6]$ & $2.613\,913\,324\,[-6]$ \\
4.7      & 60  & 1.5            & $-1.255\,937\,244\,591\,[-3]$    & $-0.317\,395\,434\,450\,53$  & $2.6\,[-2]$ & $1.763\,231\,24\,[-6]$ & $2.166\,949\,31\,[-6]$  \\
4.6      & 70  & 2.5            & $-4.261\,509\,490\,45\,[-4]$     & $-0.314\,026\,537\,276\,80$  & $2.7\,[-2]$ & $1.450\,522\,64\,[-6]$ & $1.844\,545\,73\,[-6]$  \\
\hline
\hline
\end{tabular}
\end{center}
\label{table_2p_1s_E1M1_E1E2_Yukawa}
\end{table}
\twocolumngrid

\noindent gauge invariance $\Delta_{l-v}<10^{-15}$ is found for all screening lengths.

For the $E1M1$ contribution, the $2s_{1/2}$ state is an allowed intermediate state. In the Coulomb case, $2s_{1/2}$ and $2p_{1/2}$ states are degenerate, which is not true for Yukawa potentials with $\mu \neq 0$, where $2p_{1/2}$ lies higher than $2s_{1/2}$ in the energy spectrum. Their energy difference increases with $\mu$. As $2s_{1/2}$ lies between the upper $2p_{1/2}$ and the lower $1s_{1/2}$ states of the transition, the denominator $E_{2s_{1/2}}-E_{2p_{1/2}}+\omega_1$ in \Eq{eq_Sj21} vanishes for a photon energy $\omega_1^R=E_{2p_{1/2}}-E_{2s_{1/2}}$. A sharp peak will thus appear near this resonance energy. \tablename{~\ref{table_2p_1s_E1M1_E1E2_Yukawa}} lists the non-resonant contribution to the $E1M1$ decay rate as well as the $E1E2$ $2p_{1/2} \rightarrow 1s_{1/2}$ decay rate for Yukawa potentials with $V_0=1$ and screening lengths $\delta=1/\mu$ (in a.u.). The non-resonant contribution, denoted as $E1M1^\ast$ in the table, corresponds to the area under the curve of the $E1M1$ differential decay rate, neglecting the resonance near $\omega_1^R=E_{2p_{1/2}}-E_{2s_{1/2}}$. Because of the sharp peaks near the resonance frequencies, the integration over the photon energy $\omega_1$ in \Eq{eq_WTh1L1Th2L2} requires the use of very efficient techniques to treat resonances, as detailed in \Ref{ASP09}. We do not present such calculations, since the decay of the $2p_{1/2}$ state towards the $1s_{1/2}$ ground state is largely dominated by an allowed $E1$ transition.

Similarly to the $2E1$ $2s_{1/2} \rightarrow 1s_{1/2}$ transition, the decay rates decrease with the screening length $\delta$. For $\delta<4.6$ a.u. the $2p_{1/2}$ state is not bound any more. This value is higher than for the $2E1$ $2s_{1/2} \rightarrow 1s_{1/2}$ transition. The resonance energy, denoted as $\omega_1^R$ in \tablename{~\ref{table_2p_1s_E1M1_E1E2_Yukawa}}, increases when $\delta$ decreases. It is shown by the values of $y_1^R=\omega_1^R/\omega_t$, which represents the fraction of the transition energy $\omega_t=E_{2p_{1/2}}-E_{1s_{1/2}}$ carried by the first photon at the resonance. However, the values of $\omega_1^R$ remain very small in comparison with $\omega_t$, leading to very small values of $y_1^R$. This is illustrated in \figurename{~\ref{fig_E1M1_Yukawa}}, which shows the spectral distribution function $\psi(y,V_0=1)$ of the $E1M1$ $2p_{1/2} \rightarrow 1s_{1/2}$ transition for the Coulomb potential (solid line) and a Yukawa potential with $\mu=1/5$ a.u. (dashed-dotted line). Here $n$ is equal to 8 in \Eq{eq_spectral_distr}.
\begin{figure}[ht!]
\begin{center}
\includegraphics[scale=0.4725]{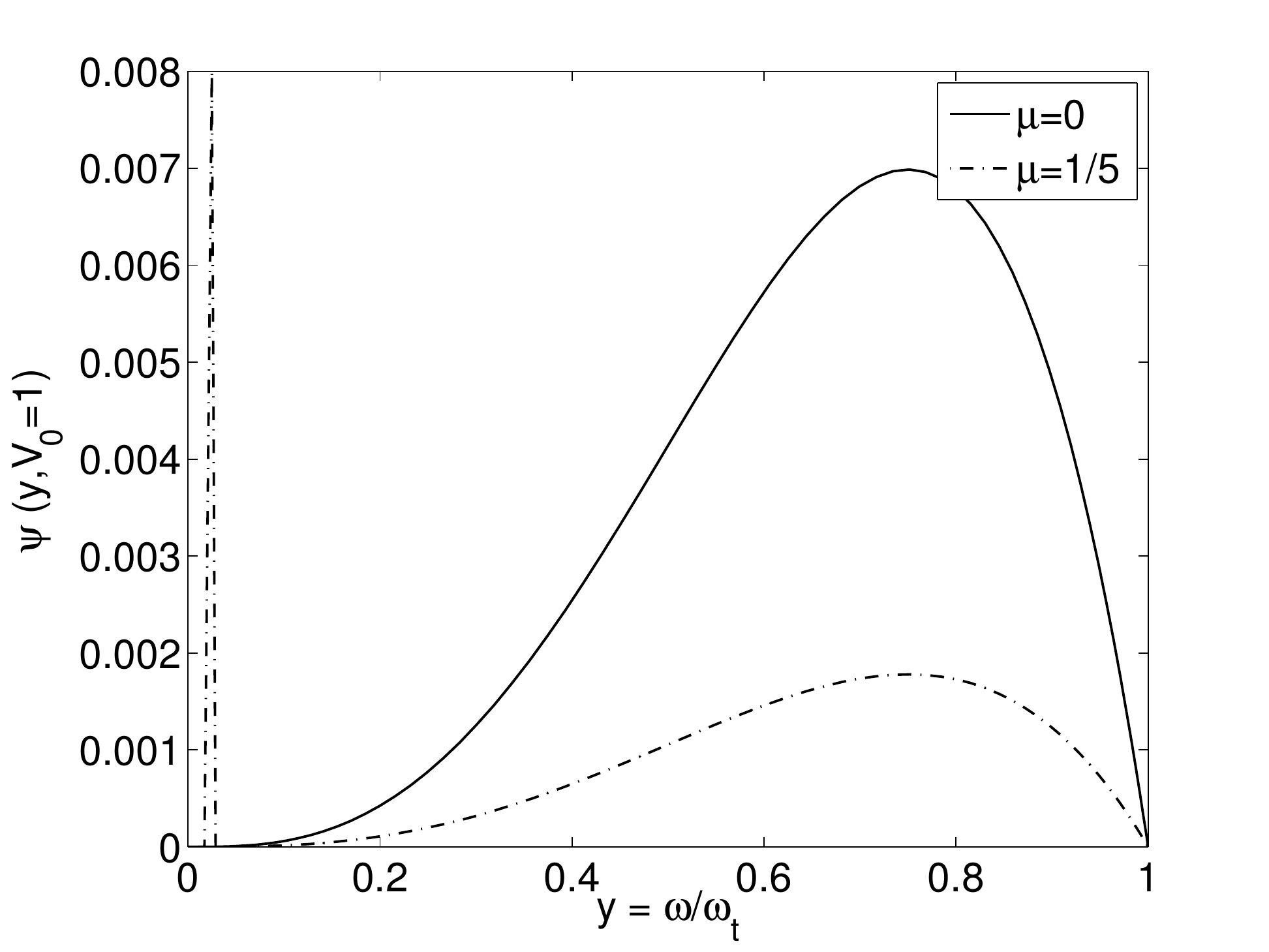}
\end{center}
\caption{{\footnotesize Spectral distribution function $\psi(y,V_0=1)$ of the $E1M1$ $2p_{1/2} \rightarrow 1s_{1/2}$ transition for the Coulomb potential (solid line) and a Yukawa potential with $V_0=1$ and $\mu=1/5$ a.u. (dashed-dotted line). The variable $y=\omega/\omega_t$ is the fraction of the photon energy carried by the first photon. The resonance occurs at $y^R=0.025$.}}
\label{fig_E1M1_Yukawa}
\end{figure}
As the resonance lies at very low frequencies $\omega_1$, the multiplication by the product $\omega_1 \omega_2$ in \Eq{eq_dWL1lam1L2lam2} will reduce the contribution of the resonance to the total decay rate. A similar effect is discussed in \Ref{CS06}, considering the $2s_{1/2} \rightarrow 1s_{1/2}$ transition in hydrogenic ions, taking into account the Lamb shift between the $2p_{1/2}$ and $2s_{1/2}$ states. The $2p_{1/2}$ level lying under the $2s_{1/2}$ level, a sharp peak appears in \Eq{eq_dWL1lam1L2lam2} near the resonance energy $\omega_1^R=E_{2s_{1/2}}-E_{2p_{1/2}}$. Increasing $Z$ increases this shift. In this reference, the authors argue that such a resonance should not contribute beyond the percent level to the total lifetime of the $2s_{1/2}$ state, as measurements for hydrogenic He and Ar show. This is also expected because the lifetime of the $2s_{1/2}$ state should not be strongly altered by the slow $2s_{1/2} \rightarrow 2p_{1/2}$ transition ($\sim 1.6 \times 10^{-9}$ s$^{-1}$ for H)~\cite{CS06}.

As $E1M1$ involves two photons with different characteristics, one expects $\psi(y,V_0=1)$ not to be symmetric around $y=0.5$, as shown in \figurename{~\ref{fig_E1M1_Yukawa}}. This argument is also valid for $M1E1$, but the sum $E1M1+M1E1$ is symmetric around $y=0.5$. \figurename{~\ref{fig_E1E2_Yukawa}} is the equivalent of \figurename{~\ref{fig_E1M1_Yukawa}} for the $E1E2$ contribution. The argument of symmetry is also valid, i.e., the sum $E1E2+E2E1$ is symmetric around $y=0.5$. From the Coulomb case $\mu=0$ to the case $\mu=1/4.6$ a.u. the maximum $\psi(0.41,V_0=1)$ decreases from $4.6 \times 10^{-3}$ to $1.2 \times 10^{-3}$.
\begin{figure}[ht!]
\begin{center}
\includegraphics[scale=0.4725]{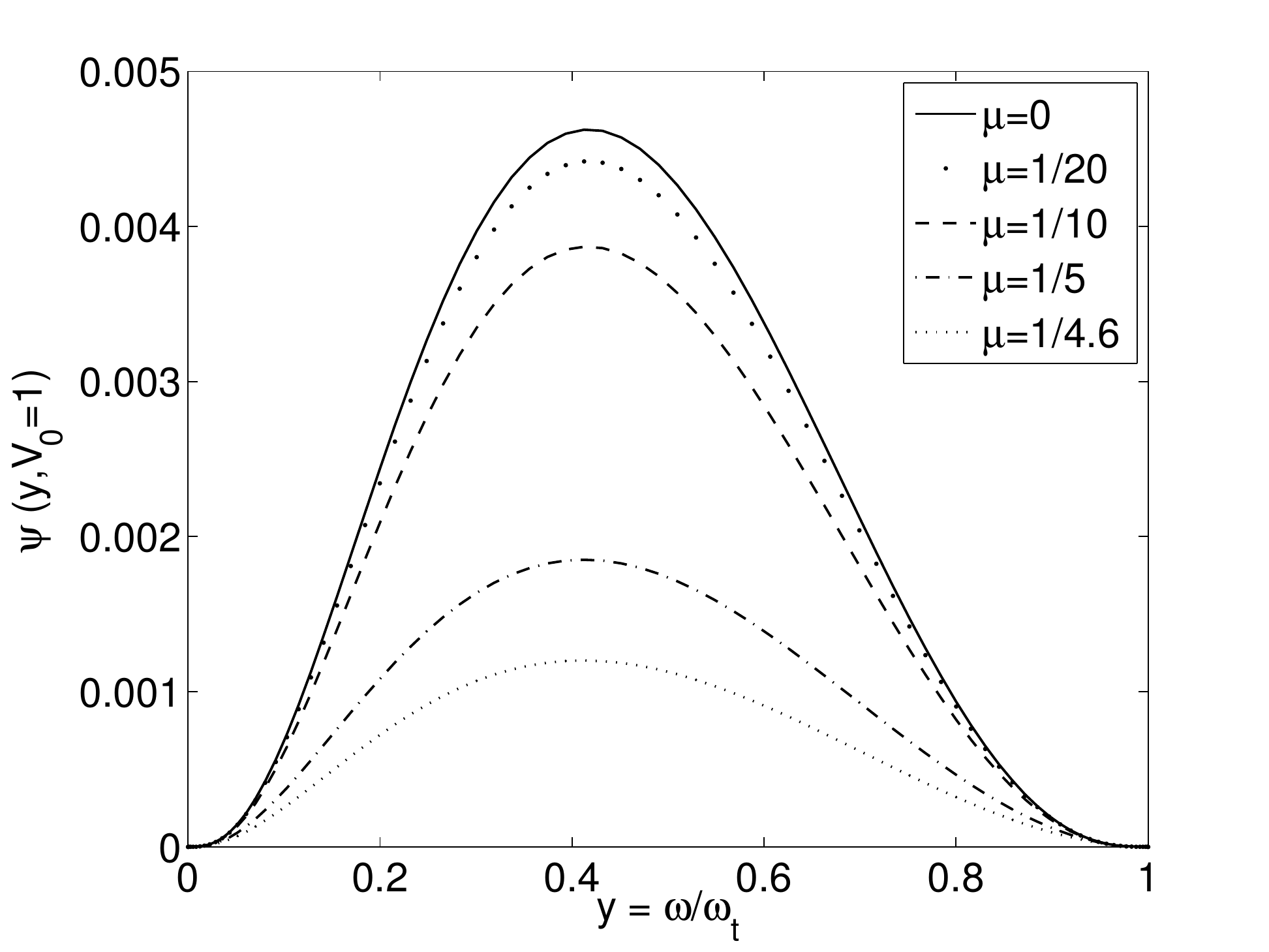}
\end{center}
\caption{{\footnotesize Spectral distribution function $\psi(y,V_0=1)$ of the $E1E2$ $2p_{1/2} \rightarrow 1s_{1/2}$ transition for Yukawa potentials with $V_0=1$ and given $\mu$ values (in a.u.). The variable $y=\omega/\omega_t$ is the fraction of the photon energy carried by the first photon.}}
\label{fig_E1E2_Yukawa}
\end{figure}

Unlike $E1M1$, the $E1E2$ contribution is non-resonant because the $2s_{1/2}$ is not a part of the allowed intermediate states for this transition. Hence, no resonance occurs in \figurename{~\ref{fig_E1E2_Yukawa}}.

%%%%%%%%%%%%%%%%%%%%%%%%%%%%%%%%%%%%%%%%%%%%%%%%%%%%%%%%%%%%%%%%%%%%%%%%%%%%%%%%%%%%%%%%%%%%%%%%%%%%%%%%%%%%%%%%%%%%%%%%%%%%%%%%%%%%%%%%%%%%%%%%%%%%%%%%%%%%%%%%%%%%%%%%%%%%%%%%%%%%%%%%%%%%%%%%%%%%%%%%%%%%%%%%%%%%%%%%%%%%%%%%%%%%%%%%%%%%%%%%%%%%
%%%%%%%%%%%%%%%%%%%%%%%%%%%%%%%%%%%%%%%%%%%%%%%%%%%%%%%%%%%%%%%%%%%%%%%%%%%%%%%%%%%%%%%%%%%%%%%%%%%%%%%%%%%%%%%%%%%%%%%%%%%%%%%%%%%%%%%%%%%%%%%%%%%%%%%%%%%%%%%%%%%%%%%%%%%%%%%%%%%%%%%%%%%%%%%%%%%%%%%%%%%%%%%%%%%%%%%%%%%%%%%%%%%%%%%%%%%%%%%%%%%%

\section{Conclusion}
\label{sec:conc}

The Lagrange-mesh method is able to provide numerically exact energies and wave functions for the Coulomb-Dirac problem. As shown in \Ref{BFG14}, some matrix elements are exactly given by the associated Gauss quadrature. For relativistic multipolar polarizabilities, \Ref{FGB14} devised a simple calculation involving different meshes for the initial and final wave functions and for the calculation of matrix elements. This calculation provides very accurate values for all charges $Z$, for the ground state and excited states of hydrogenic atoms with Coulomb or Yukawa potentials. In the present work, it is generalized to the determination of relativistic two-photon decay rates.

Similarly to the polarizabilities, the evaluation of two-photon atomic transitions is performed within the framework of the second-order perturbation theory, involving an infinite number of intermediate virtual states simulated by a finite number of pseudostates. The simplicity of the Lagrange-mesh method allows a simple extension of the technique used for polarizabilities to the relativistic $2s_{1/2} \rightarrow 1s_{1/2}$ and $2p_{1/2} \rightarrow 1s_{1/2}$ two-photon decay rates in hydrogenic atoms. Very accurate values of these rates are obtained for all charges $Z$, with a simple code and small computing times. The general requirement of gauge invariance is successfully tested, which emphasizes the high accuracy of the Lagrange-mesh method. The results with a Coulomb potential perfectly agree with benchmark values presented in Refs.~\cite{ASP09},~\cite{ASP11} and~\cite{BTE14}. They are even more accurate, providing for some partial rates up to five more significant digits than in Refs.~\cite{ASP09} and~\cite{BTE14}. This work also improves the values from \Ref{ASP11} by using a more precise value for the atomic unit of time. For $Z = 1$ to 100, total $2s_{1/2} \rightarrow 1s_{1/2}$ decay rates are obtained with eight significant digits.

The present approach is also valid for other potentials, with or without a singularity at the origin. Its efficiency and simplicity are illustrated with Yukawa potentials which simulate a hydrogen atom embedded in a Debye plasma. We studied the influence of the screening length $\delta$ on the value of the $2s_{1/2} \rightarrow 1s_{1/2}$ and $2p_{1/2} \rightarrow 1s_{1/2}$ decay rates. The plasma density scaling as $1/\delta^3$, the two-photon decay rates decrease as the plasma becomes denser. Excellent accuracies and gauge invariances of the results are also obtained for both transitions with Yukawa potentials. Properties of alkali-like atoms can easily be estimated by combining the present approach with the use of model and parametric potentials \cite{Jo07}.

The Lagrange-mesh method is definitely accurate for estimating relativistic two-photon decay rates of hydrogenic systems. As such, Lagrange bases can be used to investigate other properties of two-photon transitions, such as resonances~\cite{ASP09} and negative-continuum \cite{SSA09} effects, or more recently angular correlation and degree of linear polarization~\cite{AFF12}.

Lagrange functions, without the associated Gauss quadrature or with a partial use of this quadrature limited to terms for which it is accurate, promise to be very efficient for the study of many-electron systems in atomic physics. This basis could offer interesting simplifications in comparison with, for instance, $B$~splines confined to a large cavity~\cite{SJ96,SJ08} which have had a tremendous impact in atomic many-body calculations. The replacement of $B$~splines by a Sturmian basis~\cite{Sz97,TKS10,BTE14} in relativistic calculations of atomic properties~\cite{SSK07} is presently considered~\cite{Sa14}. From this respect, we would like to point out that specific Lagrange-Laguerre bases are exactly equivalent to Sturmian bases (see Appendix~\ref{sec:sturmians}), but can be simpler to use and more flexible.

%%%%%%%%%%%%%%%%%%%%%%%%%%%%%%%%%%%%%%%%%%%%%%%%%%%%%%%%%%%%%%%%%%%%%%%%%%%%%%%%%%%%%%%%%%%%%%%%%%%%%%%%%%%%%%%%%%%%%%%%%%%%%%%%%%%%%%%%%%%%%%%%%%%%%%%%%%%%%%%%%%%%%%%%%%%%%%%%%%%%%%%%%%%%%%%%%%%%%%%%%%%%%%%%%%%%%%%%%%%%%%%%%%%%%%%%%%%%%%%%%%%%
%%%%%%%%%%%%%%%%%%%%%%%%%%%%%%%%%%%%%%%%%%%%%%%%%%%%%%%%%%%%%%%%%%%%%%%%%%%%%%%%%%%%%%%%%%%%%%%%%%%%%%%%%%%%%%%%%%%%%%%%%%%%%%%%%%%%%%%%%%%%%%%%%%%%%%%%%%%%%%%%%%%%%%%%%%%%%%%%%%%%%%%%%%%%%%%%%%%%%%%%%%%%%%%%%%%%%%%%%%%%%%%%%%%%%%%%%%%%%%%%%%%%

\begin{acknowledgments}
This text presents research results of the interuniversity attraction pole programme P7/12 initiated by the Belgian-state Federal Services for Scientific, Technical and Cultural Affairs. We would like to thank P. Amaro for a private communication about his calculations. L.F. acknowledges the support from the FRIA.
\end{acknowledgments}

%%%%%%%%%%%%%%%%%%%%%%%%%%%%%%%%%%%%%%%%%%%%%%%%%%%%%%%%%%%%%%%%%%%%%%%%%%%%%%%%%%%%%%%%%%%%%%%%%%%%%%%%%%%%%%%%%%%%%%%%%%%%%%%%%%%%%%%%%%%%%%%%%%%%%%%%%%%%%%%%%%%%%%%%%%%%%%%%%%%%%%%%%%%%%%%%%%%%%%%%%%%%%%%%%%%%%%%%%%%%%%%%%%%%%%%%%%%%%%%%%%%%
%%%%%%%%%%%%%%%%%%%%%%%%%%%%%%%%%%%%%%%%%%%%%%%%%%%%%%%%%%%%%%%%%%%%%%%%%%%%%%%%%%%%%%%%%%%%%%%%%%%%%%%%%%%%%%%%%%%%%%%%%%%%%%%%%%%%%%%%%%%%%%%%%%%%%%%%%%%%%%%%%%%%%%%%%%%%%%%%%%%%%%%%%%%%%%%%%%%%%%%%%%%%%%%%%%%%%%%%%%%%%%%%%%%%%%%%%%%%%%%%%%%%

\appendix

\section{Exact variational treatment of the Dirac-Coulomb problem with a Lagrange basis}
\label{sec:var_exact}

For the Dirac-Coulomb problem, the coupled radial Dirac equations are given by \Eq{r.1} in matrix form, where the Hamiltonian matrix $H_\kappa$ is given by \Eq{r.2} with the Coulomb potential $V(r)=-Z/r$. The radial functions $P_{n\kappa}(r)$ and $Q_{n\kappa}(r)$ are expanded in regularized scaled Lagrange-Laguerre functions \rref{Lag.3}, with $\alpha=2\gamma-2$,
\beq
\hat{f}^{(2\gamma-2)}_j (r/h) \propto \frac{L_N^{2\gamma-2}(r/h)}{r-hx_j}\, r^{\gamma} e^{-r/2h}
\eeqn{A.1}
as
\beq
\left( \begin{array}{c} P_{\kappa}(r) \\ Q_{\kappa}(r) \end{array} \right) = h^{-1/2} \left( \begin{array}{c} \sum_{j=1}^N p_{\kappa j} \hat{f}_j^{(2\gamma-2)}(r/h) \\ \sum_{j=1}^N q_{\kappa j} \hat{f}_j^{(2\gamma-2)}(r/h) \end{array} \right)
\eeqn{A.2}
with $\gamma=\sqrt{\kappa^2-\alpha^2 Z^2}$ and $L_N^{2\gamma-2}(x_i)=0$.

The coupled radial equations \rref{r.1} take the form of a $2N \times 2N$ algebraic system
\beq
\ve{H}_\kappa^G \left( \begin{array}{c} \ve{p} \\ \ve{q} \end{array} \right) = E \left( \begin{array}{c} \ve{p} \\ \ve{q} \end{array} \right)
\eeqn{A.3}
with $\ve{p}^T=(p_1,p_2,\dots,p_N)$, $\ve{q}^T=(q_1,q_2,\dots,q_N)$, and $\ve{H}_\kappa^G$ given by \Eq{Lag.7}, with a $2 \times 2$ block structure. The matrix elements of $d/dx$ are given by \Eq{Lag.8} at the Gauss quadrature approximation.

While the matrix elements of $1/r$ between regularized Lagrange functions are exactly given by the Gauss quadrature, the matrix equations \rref{A.3} are not exactly variational for two reasons. (i) As mentioned in Sec.~\ref{sec:LMM}, the Lagrange functions are not orthonormal~\cite{Ba15},
\beq
\langle \hat{f}_i^{(2\gamma-2)} \vert \hat{f}_j^{(2\gamma-2)} \rangle = \delta_{ij} + \frac{(-1)^{i-j}}{\sqrt{x_i x_j}}.
\eeqn{A.4}
(ii) The antisymmetric exact matrix representing $d/dx$ reads
\beq
\ve{D} = \ve{D}^G - \frac{1}{2} \ve{v} \ve{v}^T
\eeqn{A.5}
where $\ve{v}^T=(v_1,v_2,\dots,v_N)$ with
\beq
v_i = \frac{(-1)^i}{\sqrt{x_i}}.
\eeqn{A.6}

The exact variational equations can be written as
\beq
& & \left[ \ve{H}_\kappa^G + \left( \begin{array}{c c} 0 & -\frac{c}{2h} \ve{v} \ve{v}^T \\ -\frac{c}{2h} \ve{v} \ve{v}^T & -2c^2 \ve{v} \ve{v}^T \end{array} \right) \right] \left( \begin{array}{c} \ve{p} \\ \ve{q} \end{array} \right) \eol
& & = E \left[ I + \left( \begin{array}{c c} \ve{v} \ve{v}^T & 0 \\ 0 & \ve{v} \ve{v}^T \end{array} \right) \right] \left( \begin{array}{c} \ve{p} \\ \ve{q} \end{array} \right)
\eeqn{A.7}
where $I$ is the $2N \times 2N$ identity matrix. This is thus a generalized eigenvalue problem, contrary to \Eq{A.3}.

In fact, for the scaling parameter $h=\mathcal{N}/2Z$, the variational equations \rref{A.7} provide the exact solution of the Dirac-Coulomb problem because this solution can be exactly expressed as a combination of the Lagrange functions \rref{A.1}. Moreover, as shown below, Eqs. \rref{A.3} and \rref{A.7} have exactly the same exact eigenvalue and eigenvector and the Lagrange-mesh equations \rref{A.3} also provide the exact solution of the Dirac-Coulomb problem.

This property is due to the fact that the exact solutions of \Eq{r.1} are polynomials times the square root of the Laguerre weight function used here~\cite{Ba15}. Indeed, the following integral involving the large component $P_{n\kappa}$ exactly vanishes for $n<N$,
\beq
\int_0^\infty r^{-1} P_{n\kappa}(r) L_{N-1}^{2\gamma-2}(r/h) r^{\gamma-1} e^{-r/2h} \, dr = 0.
\eeqn{A.8}
With the change of variable $r=hx$, its integrand is the product of the Laguerre weight function $x^{2\gamma-2}e^{-x}$ by the Laguerre polynomial $L_{N-1}^{2\gamma-2}(x)$ and a polynomial of degree $n-1<N-1$. These polynomials are thus exactly orthogonal. This integral can be
exactly evaluated with the corresponding Gauss quadrature as
\beq
\sum_{k=1}^N \lambda_k (hx_k)^{\gamma-2} P_{n\kappa}(hx_k) L_{N-1}^{2\gamma-2}(x_k) e^{-x_k/2} = 0.
\eeqn{A.9}
By using the relation
\beq
p_{n\kappa k} = h^{1/2} \lambda_k^{1/2} P_{n\kappa}(hx_k)
\eeqn{A.10}
derived from \Eq{A.2} with the Lagrange property $\hat{f}^{(2\gamma-2)}_j (x_k)=\lambda_k^{-1/2} \delta_{jk}$, and the square root of the Gauss weight~\cite{AS65}
\beq
\lambda_k = \dfrac{\Gamma(N+2\gamma-2) e^{x_k}}{N!(N+2\gamma-2) x_k^{2\gamma-3} [L_{N-1}^{2\gamma-2}(x_k)]^2},
\eeqn{A.11}
one obtains for $N>n$,
\beq
\sum_{k=1}^N p_{n\kappa k} (-1)^k x_k^{-1/2} = 0.
\eeqn{A.12}
More compactly, one has
\beq
\ve{v}^T \ve{p}_{n\kappa} = 0.
\eeqn{A.13}
The same property holds for the small component $Q_{n\kappa}$, i.e., $\ve{v}^T \ve{q}_{n\kappa}=0$. Hence, since $(\ve{p}_{n\kappa},\ve{q}_{n\kappa})$ is an exact solution of the variational equations \rref{A.7}, it is also an exact solution of the Lagrange-mesh equations \rref{A.3}.

%%%%%%%%%%%%%%%%%%%%%%%%%%%%%%%%%%%%%%%%%%%%%%%%%%%%%%%%%%%%%%%%%%%%%%%%%%%%%%%%%%%%%%%%%%%%%%%%%%%%%%%%%%%%%%%%%%%%%%%%%%%%%%%%%%%%%%%%%%%%%%%%%%%%%%%%%%%%%%%%%%%%%%%%%%%%%%%%%%%%%%%%%%%%%%%%%%%%%%%%%%%%%%%%%%%%%%%%%%%%%%%%%%%%%%%%%%%%%%%%%%%%

\section{Equivalence between Dirac-Coulomb Sturmians and Lagrange-Laguerre basis}
\label{sec:sturmians}

With the Coulomb potential $V(r)=-Z/r$, the components of the radial Dirac Sturmians are given for $n=-\infty$ to $+\infty$ by~\cite{Sz97}
\beq
S_{n\kappa}(2\lambda r) = r^\gamma e^{-\lambda r} s_n(r)
\eeqn{B.1}
and
\beq
T_{n\kappa}(2\lambda r) = r^\gamma e^{-\lambda r} t_n(r),
\eeqn{B.2}
where $s_n(r)$ and $t_n(r)$ are polynomials of degree $|n|$ in $r$. The constants $\gamma$ and $\lambda$ are defined by $\gamma=\sqrt{\kappa^2-\alpha^2 Z^2}$, and 
\beq
\lambda = \sqrt{-E(2c^2+E)}/c.
\eeqn{B.3}
In \Eq{B.3}, $E$ is a fixed real parameter comprised between $-2c^2$ and $0$. 

For $|n|<N$, the Sturmian components can be expanded in $N$ Lagrange functions~\rref{A.1} with 
\beq
h = 1/2\lambda.
\eeqn{B.4}
For each $N$ value, the Lagrange functions are $N$ linearly independent polynomials of degree $N-1$, multiplied by $r^\gamma \exp(-\lambda r)$. With the Lagrange property 
\beq
\hat{f}^{(2\gamma-2)}_j (x_i) = \lambda_i^{-1/2} \delta_{ij},
\eeqn{B.5}
where the $\lambda_i$ are the Gauss weights, not to be confused with parameter $\lambda$ defined in \Eq{B.3}, one can write for $|n|<N$ the exact expansions 
\beq
S_{n\kappa}(2\lambda r) = \sum_{j=1}^N \lambda_j^{1/2} S_{n\kappa}(x_j) \hat{f}^{(2\gamma-2)}_j (2 \lambda r)
\eeqn{B.6}
and
\beq
T_{n\kappa}(2\lambda r) = \sum_{j=1}^N \lambda_j^{1/2} T_{n\kappa}(x_j) \hat{f}^{(2\gamma-2)}_j (2 \lambda r).
\eeqn{B.7}
The $n=-(N-1)$ to $N-1$ large and small components of the Sturmians can be expressed exactly with linear combinations of $N$ Lagrange-Laguerre functions. 

The situation is slightly different for vectors of Dirac radial functions. A radial Dirac vector is approximated with $2N-1$ Sturmian vectors as 
\beq
\left( \begin{array}{c} P_\kappa(r) \\ Q_\kappa(r) \end{array} \right) = \sum_{n=-(N-1)}^{N-1} c_{n\kappa} \left( \begin{array}{c} S_{n\kappa}(2\lambda r) \\ T_{n\kappa}(2\lambda r) \end{array} \right),
\eeqn{B.8}
while an expansion in Lagrange functions reads according to \Eq{A.2},
\beq
\left( \begin{array}{c} P_\kappa(r) \\ Q_\kappa(r) \end{array} \right) & = & h^{-1/2} \sum_{j=1}^N \left[ p_{\kappa j} \left( \begin{array}{c} \hat{f}_j^{(2\gamma-2)}(r/h) \\ 0 \end{array} \right) \right. \eol
& & \quad \quad \quad + \left. q_{\kappa j} \left( \begin{array}{c} 0 \\ \hat{f}_j^{(2\gamma-2)}(r/h) \end{array} \right) \right].
\eeqn{B.9}
The number of basis vectors is not exactly the same but this difference should not be significant in converged numerical calculations. This difference does not exist in the non-relativistic case.

%%%%%%%%%%%%%%%%%%%%%%%%%%%%%%%%%%%%%%%%%%%%%%%%%%%%%%%%%%%%%%%%%%%%%%%%%%%%%%%%%%%%%%%%%%%%%%%%%%%%%%%%%%%%%%%%%%%%%%%%%%%%%%%%%%%%%%%%%%%%%%%%%%%%%%%%%%%%%%%%%%%%%%%%%%%%%%%%%%%%%%%%%%%%%%%%%%%%%%%%%%%%%%%%%%%%%%%%%%%%%%%%%%%%%%%%%%%%%%%%%%%%
%%%%%%%%%%%%%%%%%%%%%%%%%%%%%%%%%%%%%%%%%%%%%%%%%%%%%%%%%%%%%%%%%%%%%%%%%%%%%%%%%%%%%%%%%%%%%%%%%%%%%%%%%%%%%%%%%%%%%%%%%%%%%%%%%%%%%%%%%%%%%%%%%%%%%%%%%%%%%%%%%%%%%%%%%%%%%%%%%%%%%%%%%%%%%%%%%%%%%%%%%%%%%%%%%%%%%%%%%%%%%%%%%%%%%%%%%%%%%%%%%%%%

\end{document}